\documentclass[12pt,preprint]{aastex}

\def\kms{\,\,\,{\rm km\,s^{-1}}}                      
\def\msol{\,{\rm M}_\odot}              
\def\gcc{\,\,\,{\rm g\,cm^{-3}}}
\def\cc{\,\,\,{\rm cm^{-3}}}

\def \mp{\widetilde{M}_{\rm peak}}
\def \kcrit{K_{\rm crit}}
\def \nbar{\bar n}
\def \rbar{\bar \rho}
\def \ncrit{{\bar n}^{crit}}

\def\ga{\,\hbox{\hbox{$ > $}\kern -0.8em \lower 1.0ex\hbox{$\sim$}}\,}
\def\la{\,\hbox{\hbox{$ < $}\kern -0.8em \lower 1.0ex\hbox{$\sim$}}\,}

\begin{document}

\title{Analytical theory for the initial mass function: II.\\
Properties of the flow.}

\author
{Patrick Hennebelle }
\affil{Laboratoire de radioastronomie, UMR CNRS 8112,\\
 \'Ecole normale sup\'erieure et Observatoire de Paris,\\
24 rue Lhomond, 75231 Paris cedex 05, France }

\and

\author{Gilles Chabrier}
\affil{\'Ecole normale sup\'erieure de Lyon,
CRAL, UMR CNRS 5574,\\ Universit\'e de Lyon, 69364 Lyon Cedex 07,  France}


\date{}


\begin{abstract}

Recently, Hennebelle and Chabrier (2008) derived an analytical theory for the mass spectrum of  non self-gravitating clumps associated with overdensities in molecular clouds and for the initial mass function of gravitationally bound prestellar cores, as produced by the turbulent collapse of the cloud.
In this companion paper, we examine the  effects of  the non-isothermality of the flow, of the turbulence forcing and of local fluctuation of the 
velocity dispersion, on the mass function. In particular, we investigate  the influence of a polytropic equation of state and of the 
effective adiabatic exponent $\gamma$ and find that it has a drastic influence on the low mass part of the IMF. 
We also consider a barotropic equation of state (i.e. a piecewise polytropic eos) that mimics the thermal behaviour of the molecular gas and  compare the prediction 
of our theory with the results of numerical simulations and with the observationally-derived IMF, 
for cloud parameters which satisfy Larson's type relations. We find that for clouds 
whose density is,  at all scales, almost an order of magnitude larger than
 the density inferred for the CO clumps in the Galaxy, a good agreement is obtained between the 
theory and the observed IMF, suggesting that star formation preferentially occurs in high density environments. We derive an analytical expression for the IMF which generalizes the expression previously obtained for the isothermal case. This easy-to-implement analytical IMF should serve as a template to compare observational or numerical results with the theory.
\end{abstract}


\section{Introduction}
Understanding star formation is one of the greatest challenges of astrophysics. Amongst many outstanding 
questions, finding out what determines the stellar initial mass function (IMF) first observationally derived by Salpeter (1955) and later on
by many others (e.g. Kroupa 2002, Chabrier 2003ab, 2005), constitutes 
a key issue. Various observational works have recently suggested that the IMF could be inherited from the conditions prevailing in 
the  molecular clouds. In particular the core mass function (CMF) has been found to have a shape very similar to the IMF 
(Motte et al. 1998, Testi \& Sargent 1998, Johnstone et al. 2000, Alv\'es et al. 2007, Nutter \& Ward-Thompson 2007, Simpson et al. 2008, 
Enoch et al. 2008), although shifted 
towards larger masses by a factor of about 2-3. Although this behaviour remains to be confirmed unambiguously, it suggests that the IMF is 
rooted in the physics of the turbulent self-gravitating gas which very likely determines the core formation.
This interpretation has been questioned by Clark et al. (2007) who point out that since the lifetime of the more massive cores 
is likely to be longer than the lifetime of the less massive ones, the core mass function should be indeed flatter in order to 
lead to the IMF. This issue is investigated in  appendix C where a simple solution is proposed.
The relationship between the CMF and the IMF has also been recently discussed by Goodwin et al. (2008) and Smith et al. (2008). 
 Theoretically, IMFs very similar 
to the one inferred from observations have been obtained from hydrodynamical numerical simulations of supersonic, isothermal 
self-gravitating gas (e.g. Klessen 2001, Bate \& Bonnell 2005). 
An analytical approach of the IMF based on such a turbulence-driven star formation picture, although ignoring the turbulent support,
 has been proposed by Padoan et al. (1997) and Padoan \& Nordlund (2002). These authors conclude that in the absence of magnetic 
field, a too stiff IMF is obtained and that one needs an at least weakly magnetized gas to  obtain a CMF/IMF with a slope close to the 
Salpeter's value,
contradicting the results of Bate \& Bonnell (2005) and Jappsen et al. (2005).

Recently Hennebelle \& Chabrier (2008, hereafter paper I), have extended the formalism 
derived in cosmology by Press \& Schechter (1974) to predict the galaxy formation mass spectrum, to the context of star formation, and have derived a new
 analytical theory of the IMF. These authors show that the IMF includes two contributions, namely a power-law  at large
 scales, and a lognormal contribution which yields a truncation at both large and small scales. 
The first contribution stems from the
 scale introduced for gravitational instability, the Jeans scale in the present context, 
while the second contribution 
arises from the statistical counting
 of overdense structures exceeding some density threshold, similar to the Press-Schechter cosmological case.
This theory  is based on the same gravo-turbulent picture as the one proposed by 
Padoan et al. (1997) but, among other differences, explicitly includes the turbulent support, emphasizing the dual role played by turbulence, which leads to the enhancement of (collapsing) overdensities on one hand and to a non-thermal support of the cores on the other hand.
The theory  predicts a 
CMF/IMF very close to the observational one, even in the hydrodynamical case, without the need to invoke magnetic field, and relates explicitly the slope of the high-mass power-law part of the IMF to the power spectrum index of turbulence.
Within the same formalism, when invoking only the density threshold condition, the mass spectrum of 
non self-gravitating CO clumps, identified as  overdensities in molecular clouds, is also derived and is found to 
be in good agreement with the mass spectrum inferred observationally (e.g. Heithausen et al. 1998), 
i.e. $dN / dM = {\cal N}(M) \propto M^{-x}$ with $x \simeq 1.6-1.8$.

In the present paper, we extend various aspects of this theory, exploring the dependence 
of the IMF upon various thermal and dynamical properties of the flow, namely   non-isothermality, 
turbulence driving and local fluctuations of the velocity dispersion. As will be seen, both 
aspects are playing a significant quantitative role although not modifying qualitatively
the main features of the theory.

The paper is organized as follows: in \S2, we summarize our theory and in \S3 we investigate the influence of 
the velocity statistics on this theory. In \S4 we consider a polytropic equation of state while in \S5 we 
study the influence of a barotropic equation of state that mimics the temperature distribution of the 
molecular gas. In \S6 we  compare the theory prediction with the results of numerical simulations and 
with  the IMF inferred by Chabrier (2003, 2005), determining the set of cloud parameters 
for which good agreement is obtained. Section 7  concludes the paper.

\section{Summary of the theory}
\label{theory}
The  theory is based on an extension of the statistical Press \& Schechter (1974, PS) formalism 
(see also Bond et al. 1991, Padmanabhan 1993, Jedamzik 1995, Audit et al. 1997), developed in cosmology, to the
 context of star formation, characterized by a non-uniform (lognormal) underlying density field. 
 The principle of the method is the following. First, the density field is smoothed at scale, $R$, using a window function
(which in our case is the sharply truncated in k-space window). Then, the mass contained in  areas which at scale, $R$, 
satisfy a  specified density  criterion  is counted. This mass is equal to the mass that will end up in structures
of mass larger than a scale dependent critical mass. Taking the derivative with respect to $R$  of the corresponding 
equation, we obtain the mass spectrum.

In our theory, non self-gravitating CO clumps are identified as local overdensities, 
$\delta = \log (\rho / \bar {\rho}) $, exceeding a simple, {\it constant} density threshold $\delta_c$, while in the case of 
gravitationally-bound prestellar cores,  the density threshold is given by the requirement that a fluctuation contains
one local (thermal or turbulent) Jeans mass.
 Whereas pure thermal support yields a CMF at large scales steeper than the Salpeter
 one, $dN/d\log\, M \propto M^{-1.35}$, turbulent dispersion leads to the correct slope, for
the index of turbulence found in observations or in numerical simulations of isothermal compressible turbulence (e.g. Kritsuk et al. 2007).
 Turbulent support (e.g. Bonnazola et al. 1987, V\'azquez-Semadeni \& Gazol 1995) enters the theory through an effective Mach number:

\begin{eqnarray}
{\cal M}_* = { 1  \over \sqrt{3} } { V_0  \over C_s}\left({\lambda_J^0 \over   1 {\rm pc} }\right) ^{ \eta}
\approx (0.8-1.0) \,\left({\lambda_J^0\over 0.1\,{\rm pc}}\right)^{\eta}\,\left({C_s\over 0.2\kms}\right)^{-1},
\label{mach_eff}
\end{eqnarray}
defined as the ratio of non-thermal vs sound speed at the mean   Jeans scale $\lambda_J^0$ (and not at the local Jeans length), whereas the usual Mach number, ${\cal M}$,
 represents the same quantity at the scale of the turbulence injection scale, $L_i$, assumed to be the characteristic size of the system:
\begin{eqnarray}
{\mathcal M}={\langle V^2 \rangle^{1/2} \over C_s},
\label{mach}
\end{eqnarray}
 Since the global Mach number, ${\cal M}$,  broadens the density PDF, it describes the trends of supersonic turbulence to promote star formation by 
creating new collapse seeds  while on the contrary, the effect described by ${\cal M}_*$  is to prevent the gravitational collapse through the turbulent
dispersion.
Here, $C_s=({kT / \mu m_H})^{1/2}\approx 0.2\,({\mu / 2.0})^{-1/2}\,({T / 10\,{\rm K}})^{1/2}\kms$ 
denotes the thermal sound speed, $\mu$ is the mean molecular weight, and $V$ is the (scale-dependent) rms velocity which obeys the Larson relation (Larson 1981):
\begin{eqnarray}
\langle V_{\rm rms}(L)^2\rangle =  V_0^2 \times \left( {L \over  1 {\rm pc}} \right) ^{2 \eta},
\label{larson}
 \end{eqnarray}
with $V_0\simeq 1 \kms$ and $\eta \simeq 0.4$-0.5.
The various symbols and notations used in the paper are defined in Table 1 of paper I.

The scale-dependence for the velocity dispersion associated with  supersonic turbulence adopted in our theory is given by (see paper I for details):

\begin{eqnarray}
\sigma^2(R) = \int ^{2 \pi/R} _{2 \pi /L_i} \widetilde{\delta}^2(k) 4 \pi k ^2 dk
 = \sigma_0^2 \left( 1 - \left( {R \over L_i} \right)^{n'-3} \right),
\label{sigma_turb}
\end{eqnarray}
with
\begin{eqnarray}
\sigma_0^2=\ln (1 + b^2 {\cal M}^2)
\label{sigma_val}
\end{eqnarray}
where $\widetilde{\delta}$ is the Fourier transform of $\delta=\log(\rho/\bar{\rho})$ and  $b$ is a constant obtained from numerical simulations\footnote{Note that in paper I, we wrote $b$ instead of $b^2$ in eqn.~(\ref{sigma_val}), but the quantitative value was consistent with the widely used value $b^2\approx 0.25$ (e.g. Padoan et al. 1997)}. In this expression, 
the Mach number ${\cal M}$ can be either a hydrodynamical or an Alfv\'enic Mach number (V\'azquez-Semadeni 1994, Padoan et al. 1997,
Passot \& V\'azquez-Semadeni 1998, Ostriker et al. 2001, Kritsuk et al. 2007).
As mentioned in paper I,
in case of supersonic turbulence,  the spectral index, $n^\prime$, of $\log (\rho)$ calculated
in  isothermal hydrodynamical and MHD simulations is found to be close
to the $n=11/3$ value obtained in incompressible turbulence for the {\it velocity} field (Beresnyak et al. 2005, Federrath et al. 2008).
In our approach, the star-forming clumps issued from these large-scale turbulent 
motions are identified with over-densities $\delta=\log (\rho / {\bar \rho})$ and
the mass associated with these over-densities  at scale $R$ is
\begin{eqnarray}
M \simeq C_m R^3 \rho = C_m  R^3 \bar{\rho}\, e^\delta,
\label{masse}
\end{eqnarray}
where $C_m$ is a geometrical coefficient of the order unity which depends on the window function,
as discussed by Lacey \& Cole (1994). For a top hat function, it may be simply equal to $4 \pi/3$ 
whereas for the sharply truncated function in k-space, these authors suggest the value $6 \pi^2$. 
This gives for   the smoothing length,   $R = (M/ (C_m \bar{\rho})) ^{1/3} \exp(-\delta/3) $.
Unlike in the PS formalism,
the mass depends not only on the spatial scale, $R$, but also on the variable $\delta$, 
a consequence of the non-uniform, lognormal density fluctuation distribution produced by turbulence.

The mass spectrum of  the number-density, $ {\cal N} (M)={d N/ dM}$, of non self-gravitating clumps
defined by a density threshold $\delta_c$ is found to be (see also Hennebelle \& Audit 2007):
\begin{eqnarray}
{\cal N}(M) = { \bar{\rho}  \over (M) ^2 } {  (n'-3) \sigma_0^2 \over 3 \sqrt{2 \pi}  \sigma ^3  } \left( { M \over M_{0}} \right)^{{n'-3\over3} }
\left( {\bar{\rho}  \over \rho_c } \right)^{{n'-3\over3}} 
 \times 
\left( \delta_c + {\sigma^2 \over 2} \right) 
\exp \left(  -{(\delta_c-{\sigma^2\over 2} )^2   \over 2 \sigma^2}  \right),  
\label{spec}
\end{eqnarray}
where $M_0=C_m \bar{\rho} L_i^3$ is the whole mass contained within a volume
$C_m L_i^3$ and $\rho_c = \bar{\rho}\, e^{\delta_c}$. 

Within the same theoretical formalism but with a density threshold which now depends on density and thus on the scale, namely the Jeans stability criterion, the IMF of self-gravitating cores is found to be:
\begin{eqnarray}
{\cal N} (\widetilde{M} ) &=& 2\, {\cal N}_0 \, { 1 \over \widetilde{R}^6} \,
{ 1 + (1 - \eta){\cal M}^2_* \widetilde{R}^{2 \eta} \over
[1 + (2 \eta + 1) {\cal M}^2_* \widetilde{R}^{2 \eta}] }
\times    \left( {\widetilde{M} \over \widetilde{R}^3}  \right) ^{-{3 \over  2} -   {1 \over 2 \sigma^2} \ln (\widetilde{M} / \widetilde{R}^3) }
\times {\exp( -\sigma^2/8 ) \over \sqrt{2 \pi} \sigma },
\label{grav_tot2}
\end{eqnarray}
where $\widetilde{R}= R / \lambda_J^0$,
$\widetilde{M} =  M / M_J^0 = \widetilde{R}\,
(1+ {\cal M}^2_* \widetilde{R}^{2 \eta})$, $\delta_R^c = \ln \bigg\{ (1 + {\cal M}^2_* \widetilde{R}^{2 \eta}) / \widetilde{R}^2  \bigg\} $, ${\cal N}_0=  \bar{\rho} / M_J^0$ and $M_J^0$, $\lambda_J^0$ denote the usual thermal Jeans mass and pseudo Jeans length\footnote{Strictly speaking the 
thermal Jeans length is $ \sqrt{\pi} C_s / \sqrt{G{\bar \rho}}$, the pseudo Jeans length differs from it by 
a constant factor.}, respectively:

\begin{eqnarray}
M_J^0&=& a_J\,{ C_s^3 \over \sqrt{G^3 \bar{\rho}}}\approx 1.0\,\, a_J \,({T \over 10\,{\rm K}})^{3/2}\,
({\mu \over 2.33})^{-1/2}\,({{\bar n} \over 10^4\,{\rm cm}^{-3}})^{-1/2}\, \msol 
\label{mjeans} \\
\lambda_J^0&=& \left( {a_J \over C_m} \right)^{1/3}{C_s\over \sqrt {G{\bar \rho}}}\approx 0.1\, a_J^{1/3} \,({T \over 10\,{\rm K}})^{1/2}\,
({\mu \over 2.33})^{-1/2}\,({{\bar n} 
\over 10^4\,{\rm cm}^{-3}})^{-1/2}\,\, {\rm pc}
\label{ljeans}
\end{eqnarray}
where $a_J$ is a dimensionless geometrical factor of order unity. Taking for example the standard definition of 
the Jeans mass, as the mass enclosed in a sphere of diameter equal to the Jeans length, we get 
$a_J=\pi^{5/2}/6$ and in this case the pseudo Jeans length is just equal to the Jeans length divided by 2.
Note, however, that the exact derivation of  $a_J$ is not very accurately determined since it is related to the criterion 
for a mass of gas to collapse. In particular, since the density distribution is not homogeneous and since 
turbulent dispersion or support also enter the calculations, no precise value can be inferred. As seen in 
\S~\ref{comp_simu}, the above mentioned estimate leads to good agreement with the simulations, even though 
a more refined calibration might be required at some stage.

The exponent
$\eta$ in eqns.~(\ref{mach_eff}) and (\ref{larson}) is related to the aforementioned (3D) index of turbulence $n$  by 
\begin{eqnarray}
 \eta=\frac{n-3}{2}.
 \end{eqnarray}
As mentioned above and in paper I, strictly speaking,
the index of turbulence $n$ which appears in this expression, which is related to the velocity power spectrum,
is not necessarily the same as the one which enters eqn.~(\ref{sigma_turb}),
$n'$, which is related to the power spectrum of $\log \rho$.  
However, as mentioned above, numerical simulations seem to find that both indexes are similar. 
 Recent high resolution simulations of non-magnetized isothermal supersonic turbulence (Kritsuk et al. 2007)
 yield $n \sim 3.8$-3.9, i.e. $\eta \sim 0.4$-0.45.

We recall (see paper I) that the present theory yields an analytical relationship between the high-mass power-law index of the
aforementioned clump and core mass functions, $d{\cal N}/dM\propto M^{-(1+x)}$, and the turbulence power spectrum indexes, such as:
\begin{eqnarray}
x&=&2-{n^\prime \over 3} {\hskip 1.cm} {\rm for\, the\, non\, self-gravitating\, clumps}\noindent \\
x&\approx& {n + 1 \over 2 n - 4}{\hskip 1.cm} {\rm for\, the\, self-gravitating\, cores},
\label{x_salp}
\end{eqnarray}
which, for the second relation, yields the Salpeter value, $1+x\simeq 2.33$ for $n\sim 3.8$, although, strictly speaking, the exact relation involves a second correcting term (see \S6.2.2 below).

 Note that, as evident from eq.~(\ref{sigma_turb}) of the present paper and eq.~(33) of paper I 
(which entails the term $d \sigma / dR$ arising  from the  calculation of $d P_R/dR$), the value $n'=3$  appears to be critical. 
If $n' \le 3$, small scale density
fluctuations are dominant while the contrary is true when $n' \ge 3$. As recalled above, tridimensional  turbulent 
flows produce powerspectra whose index at scale smaller than the injection scale, 
is typically of the order of 11/3 and therefore larger than 3.  However, driving the turbulence at small scales, would produce
powerspectra with different scale dependence and would therefore have serious 
consequences on the mass spectra since it would modify  the dependence of $\sigma$ on $R$. This is consistent with the results 
of Klessen (2001) regarding the strong influence of the driving scale on the mass spectrum.

\section{Influence of the statistics of the turbulence}

\subsection{Turbulence forcing}
\label{flow}

\begin{figure}[p]
\center{\includegraphics[angle=0,width=6in]{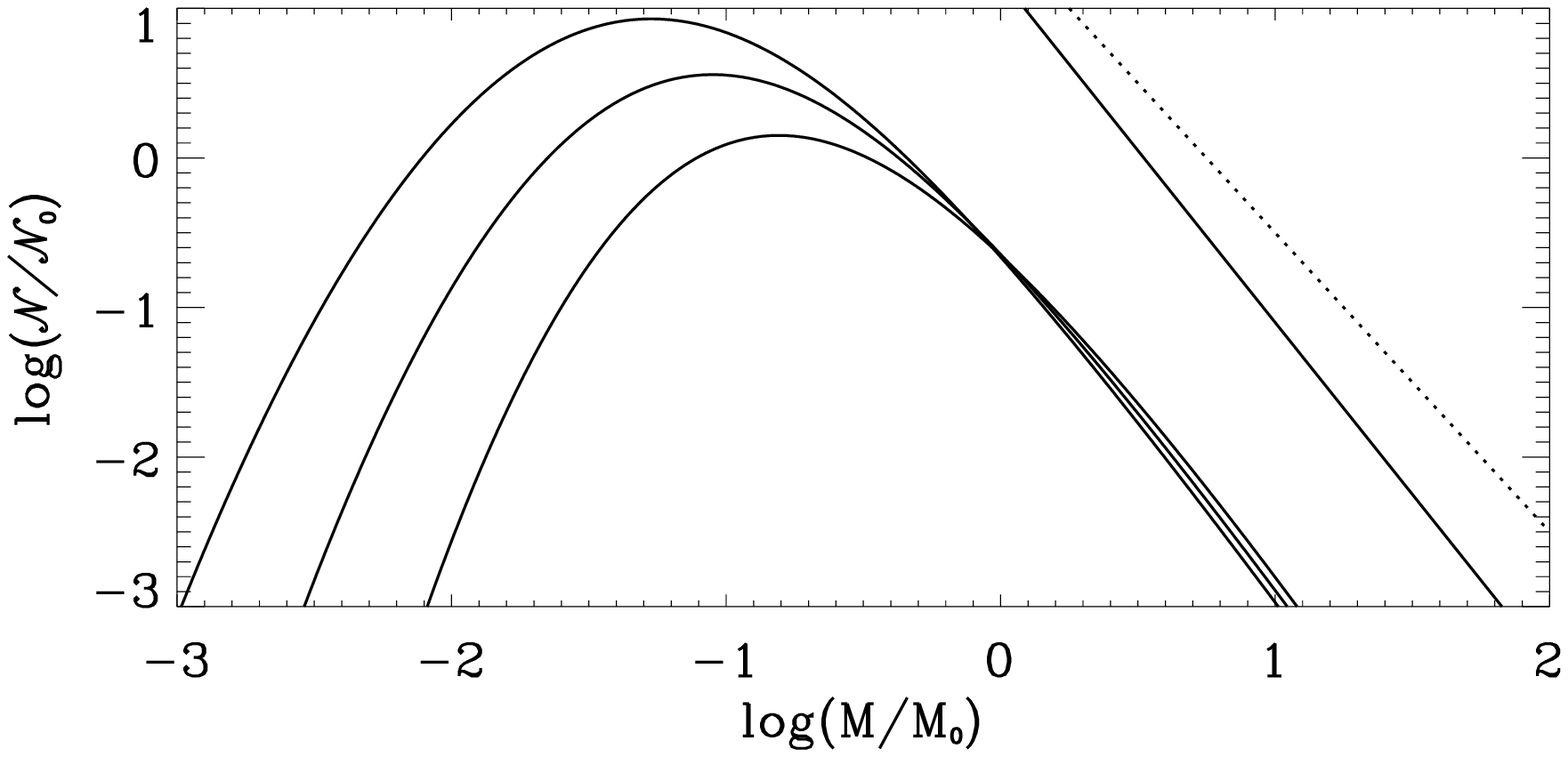}} 
\caption{Core mass function  for a Mach number ${\cal M}= 6$ and 
${\cal M}_*^2= 2$. Various values of the coefficient $b$ are shown, namely
b=0.5, 0.75 and 1 from right to left. 
 The solid and dotted lines in the upper right corner illustrate the behaviour of a Salpeter distribution,
$d{\cal N}/d \log M \propto M^{-1.35}$ and a slope $d{\cal N}/d \log M\propto M^{-1.0}$, respectively. }
\label{fig_b}
\end{figure}

The value $b\sim 0.5$ in eqn.(\ref{sigma_val}) has been derived from numerical simulations and seems to apply to both hydrodynamical
and MHD simulations (Padoan et al. 1997, Li et al. 2008). However, these simulations include only solenoidal (divergence-free) modes 
of turbulence forcing.
  Simulations including the contribution of the compressive (curl-free)
 modes in the forcing  (Federrath et al. 2008, Schmidt et al. 2009) lead to a
value of about $b\simeq1$ for the purely compressible forcing case, $b\simeq1/3$ for the 3D case with only solenoidal forcing, and values between $\sim 1/3$ and 1 when both types of modes 
contribute, as expected for the turbulence taking place in the interstellar medium. It should be noted, however, that Federrath et al. infer
PDFs that are non perfectly Gaussian, particularly at high density. Investigating the influence of such a distribution remains 
an open issue.  Figure~\ref{fig_b} portrays the theoretical CMF/IMF (eqn.(\ref{grav_tot2}))
 for $b=0.5$, $b=0.75$ and $b=1$. 
While the first value corresponds to about equipartition of solenoidal and compressive modes, 2:1 in 3D 
(equation 5 of Federrath et al. 2008 with $\zeta=1/3$), the intermediate one corresponds roughly to equipartition between total solenoidal and compressive energy, i.e. the 
compressive modes have twice as much energy as the solenoidal modes.
Not surprisingly, including compressive forcing yields a broader PDF for turbulence and thus increases the number of small-scale overdense
 regions. This shows that, as mentioned by Federrath et al. (2008), the statistics (PDF) generated by compressible turbulence depends not
 only on the Mach number but also on the forcing and on the relative contributions of rotationally and compressibly driven modes. 
This double dependence enters through the values of  $\cal M$ and $b$ in eqn.(\ref{sigma_val}) in the present theory.
Indeed, since $\sigma ^2 = {\rm ln} (1+b^2 {\cal M}^2)$, changing $b$ is equivalent to changing 
${\cal  M}$ in ${b} {\cal  M}$.
Thus, $b=1$ is equivalent to the  $b=0.5$ case but for a Mach number twice smaller. Since, as mentioned in paper I (fig. 5), 
good agreement between the present theory with $b=0.5$ and the Chabrier's IMF requires Mach numbers of about 12 (mostly for the low mass end of 
the IMF), this implies that a Mach number of $\simeq 6$ will be sufficient if $b \simeq 1$ to get a good agreement. Except if  otherwise specified, the calculations in the following sections have been conducted with a value $b=0.5$.

To end this section, we want to stress the following issue. Remembering that the parameter  ${\cal M}_*$ represents
the turbulent support at the Jeans length, it is likely that, when considering flows dominated by compressible
motions, its value should be reduced, since in that case a significant fraction of the modes are not quenching 
the collapse but, on the opposite, may tend to promote it.

\subsection{Fluctuation of the rms velocity within the cloud}
\subsubsection{Principles}
As recalled in  \S~\ref{theory}, the present theory crucially depends on the parameter 
$ {\cal M}_*$  (eqn.~(\ref{mach_eff})) which represents the velocity dispersion 
at the Jeans length characteristic of the cloud's mean density. 
As seen from  eqn.~(\ref{grav_tot2}), a single  averaged 
value of $ {\cal M}_*$ has been used so far. In reality, one expects the velocity dispersion to 
undergo fluctuations within clouds. The dispersion of the local Mach number has been computed in
 numerical simulations (e.g. Kritsuk et al. 2007, fig.~4).
 A somehow broad distribution has been obtained showing no obvious correlation with the gas density.
The shape of this distribution appears to be complex though clearly peaked around a mean value. 
For sake of simplicity, we will assume in this work that the distribution of the rms velocity,
$P_{rms}$, is lognormal with a width that we estimate from the numerical simulations. 
We write, with $X=\ln({\cal M}_*)$:
\begin{eqnarray}
P_{rms}(X, \overline{{\ln \cal M}}_*) = {1 \over \sqrt{2 \pi \sigma_{rms}^2} } 
\exp \left( -{ \left( \overline{{\ln \cal M}}_* - X \right) ^2 \over
 2 \sigma_{rms}^2 } \right),
\label{vrms}
\end{eqnarray}
where $ \overline{\ln {\cal M}}_*$ denotes the mean value of $\ln {\cal M}_*$.
From fig.~7 of Kritsuk et al. (2007), we estimate  that $\sigma_{rms} \simeq 0.3-0.5$, although the 
distribution  is not exactly lognormal.
 Given the lack of knowledge of  $P_{rms}$, we further assume 
that $\sigma_{rms}$ does not depend on the 
scale, i.e. that the dispersion of the rms velocity is the same at all scales.

In order to obtain the mean core mass spectrum, ${\cal N}(M,{\overline{\ln \cal M}}_*)$, 
 we must now integrate  
the core mass spectrum obtained for a single value of ${\cal M}_*$, ${\cal N} (M,{\ln \cal M}_*)$ 
(eqns.~(33) and B1 of paper I and eqn.~(\ref{grav_tot2}) of this
paper),
 over the ${\cal M}_*$ distribution: 
\begin{eqnarray}
{\cal N} (M, {\overline{\ln \cal M}}_*)  = \int _{-\infty} ^{\infty} P_{rms}(X,\overline{{\ln \cal M}}_*)\,
{\cal N} (M,X) dX
\label{int_N}
\end{eqnarray}

\begin{figure}[p]
\center{\includegraphics[angle=0,width=6in]{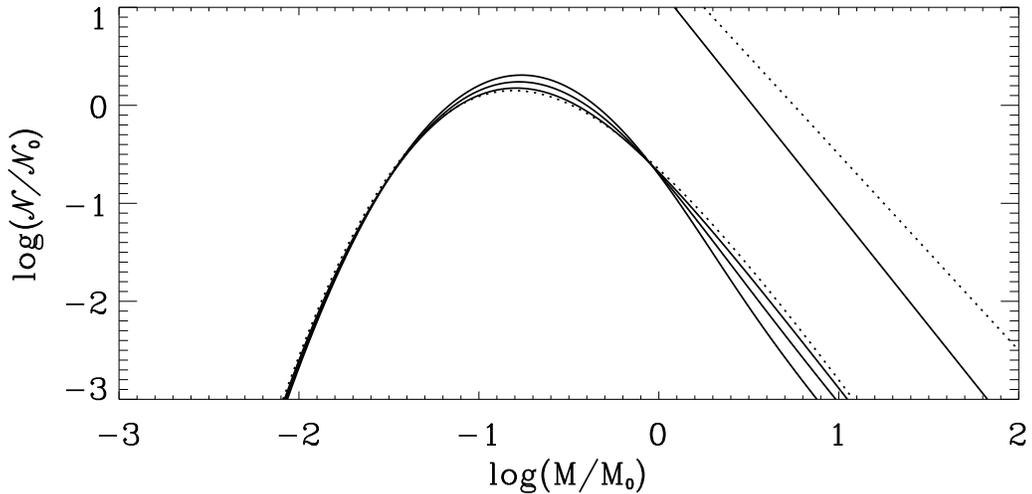}} 
\caption{Core mass spectrum for various values of $\sigma_{rms}$,  
 the width of the ${\cal M}_*$ dispersion. While the dotted line shows the 
case $\sigma_{rms}=0$, the solid lines display the cases   $\sigma_{rms}=0$, 0.2, 0.4 and 0.8.}
\label{dispersion}
\end{figure}

\subsubsection{Results}
Figure~\ref{dispersion} shows the core distribution for various values 
of $\sigma_{rms}$ and for 
 $ {\cal M} =6$ and  $ \overline{{\ln \cal M}}_* = \ln 2 / 2$.
Taking into account a distribution of ${\cal M}_*$ has a clear though modest effect on the mass spectrum by modifying the resulting core distribution
in various ways. While the amount of small-mass objects is almost unchanged, 
 more stars form in the intermediate mass range whereas less massive stars are produced, resulting in a slightly stiffer CMF/IMF with increasing $\sigma_{rms}$. This is
reminiscent of the trend displayed in fig.~1 of paper I which shows that small values
of ${\cal M}_*$ tend to produce  more intermediate mass objects and less massive stars. 
Since the core mass function obtained by taking into account the dispersion 
of ${\cal M}_*$ resembles the one obtained with a unique 
but lower value of ${\cal M}_*$, 
this implies that the low  part of the ${\cal M}_*$ distribution has more influence than the 
high  part of the distribution. 
However, since the overall effect of introducing a distribution of ${\cal M}_*$ values rather than a fix value 
appears to be modest, a constant ${\cal M}_*$ will be considered in the rest of the calculations.

\section{Polytropic equation of state}

It is well established that including a detailed treatment of the thermal properties of the gas in the molecular cloud leads 
to an equation of state, $P=K \rho^\gamma$, softer than isothermal, with $\gamma \approx 0.7$-0.8 (Larson 1985, Scalo et al. 1998, 
Glover \& Mac Low 2007). 
In this section, we explore the consequences of the departure of the gas from
isothermality on the CMF/IMF. 

\subsection{Thermal criterion with a polytropic equation of state}
We first investigate the case of a simple polytropic equation of state (eos).
The eos reads, assuming the flow behaves as a perfect gas:
\begin{eqnarray}
P=K \rho^\gamma={k\over \mu\,m_H}\rho \, T(\rho),
\end{eqnarray}
which leads for the constant $K$:
\begin{eqnarray}
K(\gamma)\equiv K ={k\over \mu\,m_H}\rho^{1-\gamma} \, T(\rho).
\end{eqnarray}
Conversely, the temperature now depends on the density as:
\begin{eqnarray}
T =T_0\,({\rho_0 \over \rho})^{1-\gamma}, 
\end{eqnarray}
where $T_0$ and $\rho_0={\bar \rho}$ denote the average initial temperature and density of the cloud.
The thermal sound speed  now depends on the local density (e.g. McKee et al. 1993, V\'azquez-Semadeni et al. 1996) as well as:

\begin{eqnarray}
C_s(\gamma)&=&\sqrt {\partial P\over \partial \rho}=C_s^0(\gamma)\,({\rho_0 \over \rho})^{{1-\gamma\over 2}}
=C_s^0(\gamma)\,e^{({\gamma -1\over 2}\,\delta)},
\label{csgam1} \\
{\rm with}\,\,\, C_s^0(\gamma)&=&\sqrt {\partial P\over \partial {\bar \rho}}=(\gamma K)^{1/2}\,{\bar \rho}^{\gamma -1\over 2}
=({k\, \gamma\over \mu m_H}\,T_0)^{1/2} \,{\bar \rho}^{1-\gamma \over 2},
\label{csgam}
\end{eqnarray}
where $\delta=\ln (\rho/{\bar \rho})$ is the density threshold for overdense structures. 

Accordingly, the mean thermal Jeans mass and pseudo Jeans length  now become:
\begin{eqnarray}
M_J^{0^\prime} (\gamma)&=& a_J\,{ (C_s^0)^3(\gamma) \over \sqrt{G^3 \bar{\rho}}}=M_J^{0}\,\gamma^{3/2}, \\
\lambda_J^{0^\prime}(\gamma)&=& \left( {a_J \over C_m} \right)^{1/3}{C_s^0(\gamma)\over \sqrt {G{\bar \rho}}}=\lambda_J^0\,\gamma^{1/2},
\nonumber
\label{lj0}
\end{eqnarray}
where $M_J^{0}$ and $\lambda_J^0$ are given by eqns.(\ref{mjeans})-(\ref{ljeans}). The local {\it thermal} Jeans mass and pseudo Jeans
length thus read:
\begin{eqnarray}
M_J&=&a_J\,{ (C_s)^3 \over \sqrt{G^3 \rho}}= M_J^{0^\prime} (\gamma)\,({\rho \over {\bar \rho}})^{{3\gamma\over 2}-2}
=M_J^{0^\prime} (\gamma)\,e^{({3\over 2}\gamma-2)\,\delta} \label{Mj}, \\
\lambda_J &=&({M_J\over \rho})^{1/3}=\lambda_J^{0^\prime}(\gamma) \,e^{({1\over 2}\gamma-1)\,\delta}.
\nonumber
\end{eqnarray}

 The conditions for {\it thermal} collapse (eqns.~(19) of paper I) now become:
\begin{eqnarray}
M\ge M_R^C &=& M_J^{0^\prime}{\tilde R}^{({3\gamma-4 \over \gamma-2})},  \\ \nonumber
\delta \ge \delta_R^C &=& {2\over \gamma -2}\ln ( {\tilde R}),
\label{crit_therm}
\end{eqnarray}
where ${\tilde R}=R/\lambda_J^{0^\prime}$.

\subsection{General criteria}
\label{crit_impl}
Proceeding as in paper I, i.e. defining an {\it effective} sound speed and Jeans mass, the  threshold condition for gravitational collapse in the general case where both thermal and non-thermal support contribute (eqns.~(28)-(29) of paper I) now reads:
\begin{eqnarray}
M\ge M_J &=& a_J { \left( C_s(\gamma)^2 + {1 \over 3} V_{rms}^2 \right)^{3/2} \over \sqrt{G^3 \rho} }.
\end{eqnarray}
With $\widetilde{R}= R / \lambda_J^{0^\prime}(\gamma)$ and
$\widetilde{M} =  M / M_J^{0^\prime}(\gamma)$,  this leads to
\begin{eqnarray}
\widetilde{M} \ge \widetilde{M}_R^c &=& \widetilde{R}  \left( \left( {\rho \over \bar{\rho} } \right)^{\gamma-1} + {\cal M}_*^2 \widetilde{R}^{2 \eta}  \right) \nonumber \\
&=& \widetilde{R}  \left( \left( { \widetilde{M} \over \widetilde{R}^3 }   \right)^{\gamma-1} + {\cal M}_*^2
\widetilde{R}^{2 \eta} \right).
\label{crit1}
\end{eqnarray}
where ${\cal M}_*^2 \equiv {\cal M}_*^2(\gamma)= \frac{1}{3}\, (V_0 / C_s^0(\gamma))^2 \times (\lambda_J^{0 \prime}/1 {\rm pc})^{2 \eta}$.
Unfortunately, eqn.~(\ref{crit1}) cannot be analytically solved and $M_R^c$ must be determined numerically:
\begin{eqnarray}
\widetilde{M}_R^c &=&  \widetilde{R}  \left( \left( { \widetilde{M}_R^c \over \widetilde{R}^3 }   \right)^{\gamma-1} + {\cal M}_*^2
\widetilde{R}^{2 \eta} \right).
\label{crit2}
\end{eqnarray}
Although resolving eqn.~(\ref{crit2}) is not difficult, 
it is convenient to have 
an analytic expression such as  eqn.~(\ref{grav_tot2}) (i.e.  eqn.~(44) of paper I), for the general $\gamma \ne 1$ case. In 
appendix A, we we derive an accurate and {\it analytic} approximation of eqn.~(\ref{crit2}) for such a general IMF. {\it This analytic expression can serve
as a quick-and-easy first approach to compare IMFs derived from either observations or numerical simulations with the present theory}.

\subsection{Core mass function with a polytropic equation of state}
Before we derive the core mass function, it is necessary to discuss the  
density PDF. As recalled in \S2, the distribution of the 
 logarithmic density fluctuations produced by isothermal turbulence  is a Gaussian with a width
given by eqns.~(\ref{sigma_turb}) and (\ref{sigma_val}).
It seems obvious from these equations that a more compressible flow ($\gamma <1$) will increase the variance of the density 
PDF for a given injection scale and a given velocity dispersion.

Indeed, the relation stated by eqn.~(\ref{sigma_val}) is valid only in the isothermal case and no 
extension has been proposed when $\gamma \ne 1$ in 3D. The density PDF itself
has not been studied and could be different from a lognormal distribution,
as discussed in Passot \& V\'azquez-Semadeni (1998) in the case of 1D flows and, though with a limited numerical resolution, 
by Li et al. (2003) in 3D.  We also recall 
that Federrath et al. (2008) find substantial departures from a lognormal distribution when 
forcing in the compressible modes, even in the isothermal case, with a significantly broader PDF. 
Since this issue appears to be unsolved yet, in this work
we will simply use eqn.~(\ref{sigma_val}) even when $\gamma \ne 1$. By doing so, we
probably, for a given Mach number, underestimate the width of the density PDF 
when $\gamma < 1$ and overestimate it when $\gamma>1$. Since, in the following we use a
fiducial Mach number ${\cal M}=6$ in our calculations, this implies that, rather than a fixed Mach number, we
{\it assume a fixed PDF width}, $\sigma$, which may correspond to  Mach numbers different from 
${\cal M}=6$ as $\gamma$ varies.
Given the lack of knowledge on the nature of turbulence in non-isothermal flows, it seems 
difficult to go further at this stage.

The mass spectrum, ${\cal N}$, is given by eqn.~(33) of paper I, leading to (neglecting the large-scale second term in the complete derivation, see \S5.1.2, 5.4 and appendix B of paper I ):
\begin{eqnarray}
{\cal N}(\widetilde{M}_R^c) = -{ \bar{\rho} \over M_J^{0 \prime} \widetilde{M}_R^c} {d \widetilde{R} \over d \widetilde{M}_R^c}  {d \delta_R^c\over d\widetilde{R}}  {1 \over \sqrt{2 \pi \sigma^2} } \exp \left( -{(\delta_R^c)^2 \over 2 \sigma^2} + 
{\delta _R^c \over 2} - {\sigma^2 \over 8}  \right), 
\label{mass_spec}
\end{eqnarray}
where 
\begin{eqnarray}
\delta_R^c = \ln \left( { \widetilde{M}_R^c \over \widetilde{R}^3} \right),
\label{dens_crit}
\end{eqnarray}

\begin{eqnarray}
{d \widetilde{M}_R^c \over d \widetilde{R} } = 
{1 \over 1 - (\gamma-1) (\widetilde{M}_R^c)^{\gamma-2} \widetilde{R}^{4-3 \gamma}  }
\left( (4-3 \gamma)  (\widetilde{M}_R^c)^{\gamma-1} \widetilde{R}^{3-3 \gamma} +
(2 \eta + 1) {\cal M}_*^2 \widetilde{R}^{2 \eta}   \right),
\label{derive_M}
\end{eqnarray}

\begin{eqnarray}
{d \delta \over d \widetilde{R} } = {d \widetilde{M}_R^c \over d \widetilde{R} } 
{1 \over  \widetilde{M}_R^c } - {3 \over  \widetilde{R} }. 
\label{derive_delt}
\end{eqnarray}

It is easily seen that for $\gamma =1$,  the isothermal limit investigated in paper I is recovered, 
while the case of pure thermal support is recovered for ${\cal M}_\star=0$.

\begin{figure}[p]
\center{\includegraphics[angle=0,width=6in]{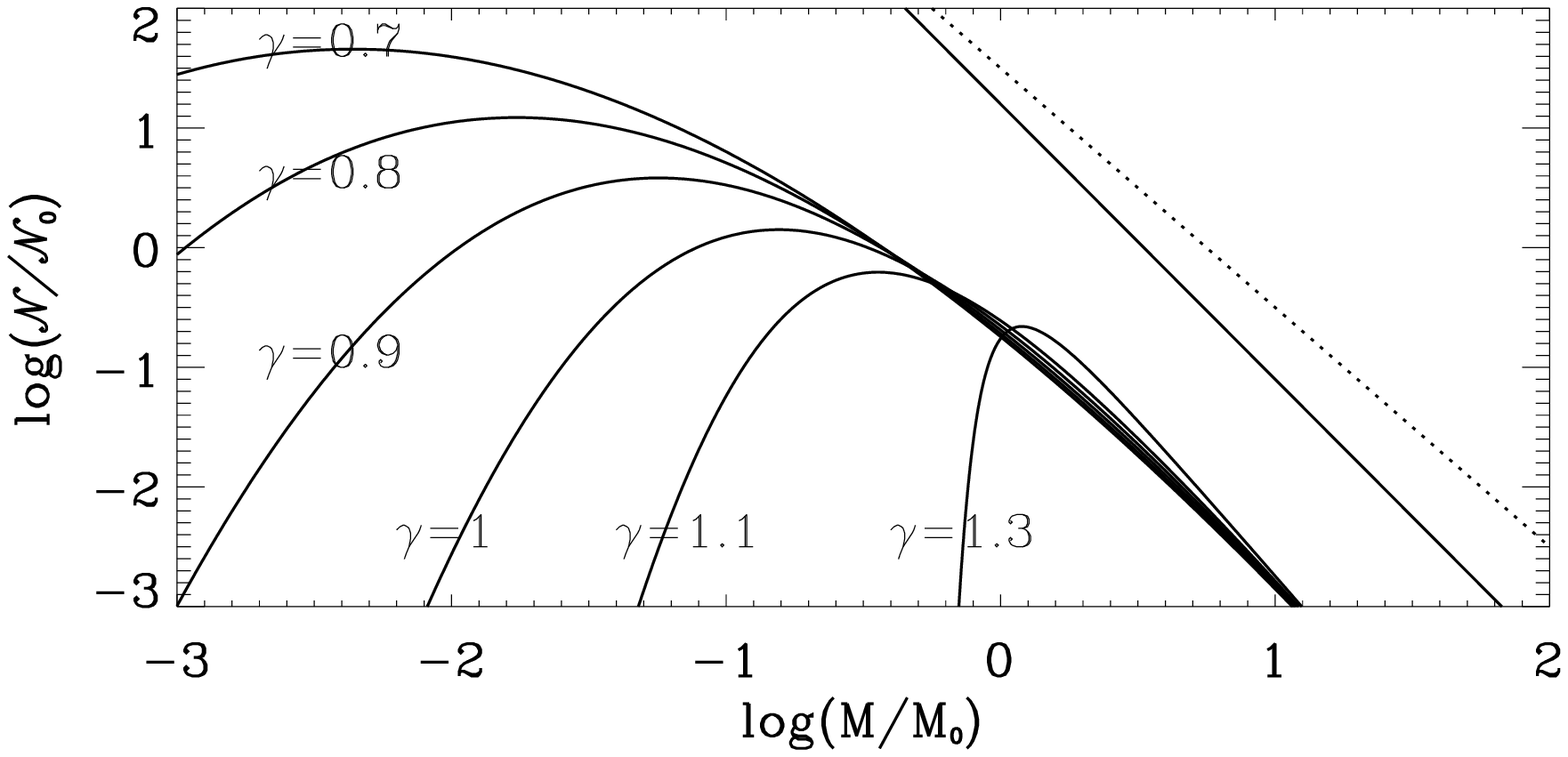}} 
\caption{Core mass function  for  ${\cal M}= 6$ and ${\cal M}_*^2= 2$ and  various values of the polytropic exponent $\gamma$, namely $\gamma=$1.3, 1.1, 1, 0.9, 0.8 and 0.7.}
\label{fig_gamma}
\end{figure}

\subsection{Peak position for a polytropic equation of state}
To obtain the peak position of the CMF, we proceed as in paper I, i.e. we derive the
core mass function, ${\cal N}$, of the purely thermal case (i.e. ${\cal M}_*=0$) 
with respect to  the mass, and calculate the mass given by  the conditions  $d {\cal N} / d M =0$. Taking into account thermal support only  is a valid 
approximation since, as shown in 
fig.~1 of paper I, ${\cal M}_*$ has a weak influence on the peak position. 

This yields:
\begin{eqnarray}
\label{minimum}
{ M_{peak} \over M_J^0 }= \mp&=& \exp \left( { (4-3 \gamma) (-9+6 \gamma) \sigma^2  \over 4} \right) \\
&=&  (1 + b^2 {\cal M}^2) ^{(3 \gamma-4) (9-6 \gamma)  \over 4}.
\nonumber
\end{eqnarray}
This equation shows that the peak position strongly varies with $\gamma$
when $1 + b^2 {\cal M}^2 \ge 1$. For example, for
$1 + b^2 {\cal M}^2 = 10$, $\gamma=1$ leads to $\mp=10^{-3/4} 
\simeq 1/6$ while $\gamma=0.7$ leads to $\mp \simeq 1/10^{2.3} 
\simeq 1/200$. A small decrease of $\gamma$ thus leads to a large variation of
the peak of the CMF. This stems for the stiff variation of the critical 
mass and density  (eqns.~(\ref{csgam}), (\ref{crit_therm}) and (\ref{dens_crit})) with $\gamma$.
As expected, when $\gamma \rightarrow 4/3$, the peak of the CMF tends 
toward $\mp=1$, i.e. $M_{peak} \rightarrow M_J^0$.

It is also interesting to compute the powerlaw index of the mass spectrum when 
turbulence is weak, (${\cal M}_* \simeq 0$). In this regime, we find that
\begin{eqnarray}
{\cal N}(M)  \propto M^{-{9 - 6 \gamma  \over 4-3 \gamma }} = M ^{-(1+x)}
\label{powerlaw}
\end{eqnarray}
This shows that the larger $\gamma$, the stiffer the core mass spectrum. 
For any reasonable value of $\gamma$,  
the value of $1+x$ remains larger than 
 $1+(n+1)/(2n-4)\simeq 2.33$ for $n\sim 3.8$ (see eqn.~(\ref{x_salp}) and paper I), a consequence of the presence of turbulent support in the latter expression. Only when $\gamma \le 0.2$, does
$1+x$ become smaller than the Salpeter's value of $\simeq 2.3$.

\subsection{Results for a polytropic equation of state}
Figure~\ref{fig_gamma} compares the CMF/IMF obtained  for various values
of $\gamma$, namely $\gamma=$1.3, 1.1, 1, 0.9, 0.8, 0.7 for a typical Mach number 
${\cal M}= 6$ and ${\cal M}_*^2= 2$ \footnote{We recall that we assume a constant 
width $\sigma$ of the lognormal distribution, which would correspond to ${\cal M}=6$ in the isothermal case.}. 
Varying $\gamma$ has a drastic influence on the core mass function.
 As expected from eqn.~(\ref{minimum}), more compressible turbulent flows ($\gamma <1$)
produce more small-scale overdense collapsing structures for a given $\sigma$
 than the isothermal gas. For $\gamma< 0.8$, the number of small objects increases dramatically.
On the contrary, the number of small mass objects decreases abruptly when $\gamma>1$.
We also see that the flow will not produce collapsing structures for $\gamma \ga 1.3$. These
results arise from the density dependence of the local Jeans mass, 
$M_J\propto \rho^{{3\over 2}(\gamma -{4\over 3})}$ (see eqn.(\ref{Mj})), which implies that the Jeans
mass {\it increases}  with density for
$\gamma > 4/3$, leading the collapse to choke itself, as pointed out in 
hydrodynamical simulations by Larson (1985), V\'azquez-Semadeni et al. (1996) and Li et al. (2003).

On the other hand, with the chosen value ${\cal M}_*^2=2$, the number of large mass 
objects is almost unchanged when $\gamma$ varies. 
This stems from the fact that, as long as $\gamma \ge 0.2$,
 turbulent support is dominant for large masses, as expected 
from eqn.~(\ref{powerlaw}).

\begin{figure}[p]
\center{\includegraphics[angle=0,width=6in]{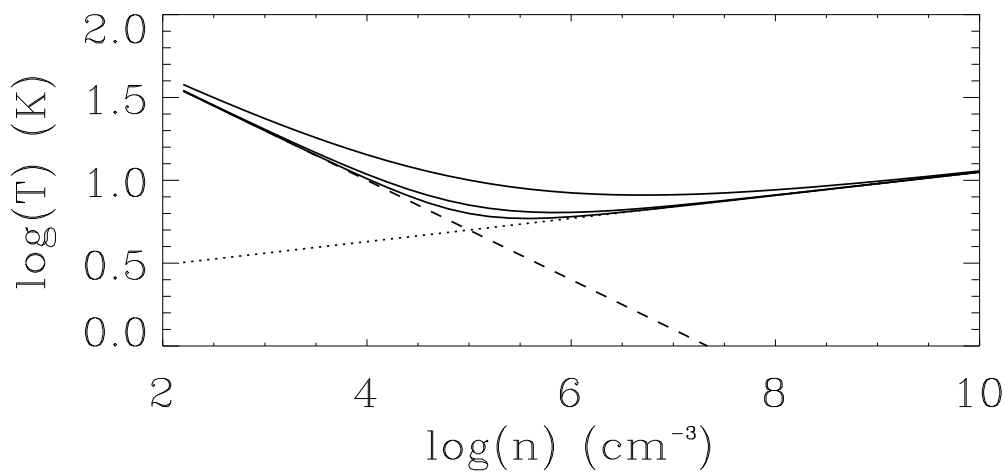}} 
\caption{Temperature distribution as stated by eqn.~(\ref{prescript}) for 
$m$ equals to  $m=1$, 2 and 3 (from top to bottom). The dotted and dashed lines correspond to 
the prescription of Jappsen et al. (2005).}
\label{temperature}
\end{figure}

\begin{figure}[p]
\center{\includegraphics[angle=0,width=6in]{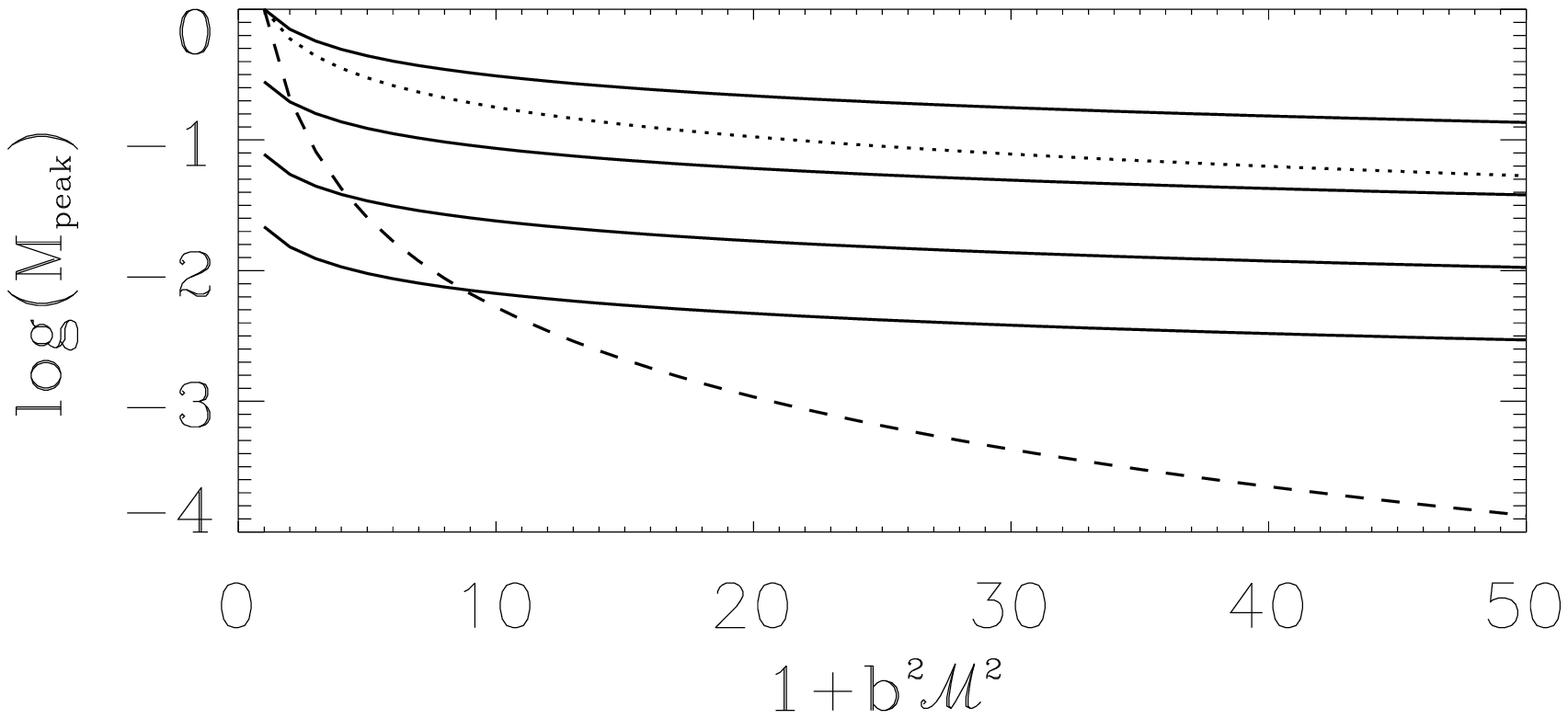}} 
\caption{Peak position of the CMF as a function of $1+b^2 {\cal M} ^2$. 
Solid lines display the peak position (eqn.~\ref{minimum_baro}) for $\gamma_2=1.1$ 
and for various values of $\rho^{crit}/{\bar \rho}=$1, 10, $10^2$ and 
$10^3$} (from top to bottom). The dashed line corresponds to the 
peak position  for $\gamma_1=0.7$ (eqn.~\ref{minimum}) whereas the dotted line 
is for $\gamma_1=1$.
\label{mass_peak}
\end{figure}

\begin{figure}[p]
\center{\includegraphics[angle=0,width=6in]{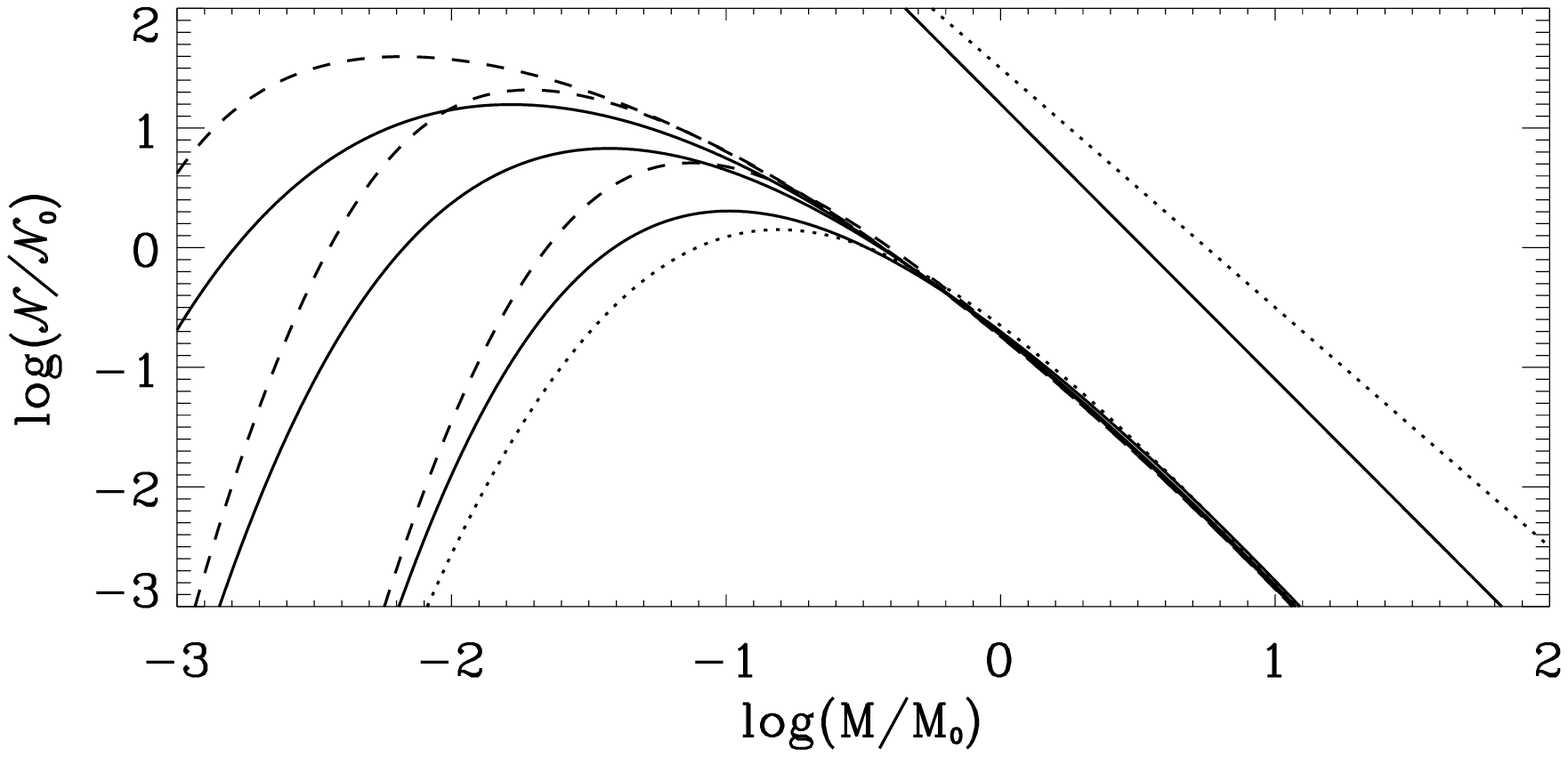}} 
\caption{Core mass function  for  ${\cal M}= 6$ and ${\cal M}_*^2= 2$, for various values of 
the constant ratio $\rho^{crit}/\bar{\rho}$, namely 10, $10^2$ and $10^3$, from right to left.
The solid lines correspond to the case $m=1$ while the dashed lines correspond 
to $m=3$.
The dotted line displays the isothermal case.}
\label{fig_gamma_var}
\end{figure}

\section{Barotropic equation of state}
\label{barotrop}
As shown in the previous section, changing $\gamma$ has a drastic influence on the mass function. 
In particular, small values of $\gamma$ tend  to produce  more small 
mass objects which form at high density in the very cold gas. However, the thermal behaviour 
of the gas in molecular clouds is not expected to be well 
described by a single powerlaw (e.g. Larson 1985). Various authors (e.g. Glover \& Mac Low 2007) found that 
for densities smaller than $\nbar \simeq 10^5$ cm$^{-3}$, i.e. $\rbar \sim 10^{-18}\gcc$, the effective polytropic exponent is 
about $\gamma \simeq 0.7$, whereas at higher densities $\gamma \simeq 1$ (Masunaga \& Inutsuka 2000).
As seen from fig.~\ref{fig_gamma},  the CMF obtained with $\gamma=0.7$ is rather different from the 
one obtained for $\gamma=1$. Thus, taking into account the variations of $\gamma$ appears to be important.

\subsection{Core mass function with a barotropic equation of state}
Jappsen et al. (2005) adopt the following prescription for the temperature variation in the clouds:
\begin{eqnarray}
\nonumber
T = a_1 \rho ^{\gamma_1-1} \; \; \rho < \rho^{crit}, \\
T = a_2 \rho ^{\gamma_2-1} \; \; \rho > \rho^{crit},  
\label{prescript_jap}
\end{eqnarray}
where the critical density, $\rho^{crit} \simeq 10^{-18}\gcc$ corresponds to ${\nbar}^{crit}=2.5 \times 10^5$ cm$^{-3}$, 
$\gamma_1=0.7$ and $\gamma_2=1.1$.
In our work, we find it more convenient to write
\begin{eqnarray}
\nonumber
C_s^2 &=& \left[  \left((C_{s,1}^0)^2 ({\rho  \over \bar{\rho} })^{\gamma_1-1}\right)^m + \left( (C_{s,2}^0)^2 ({\rho  \over \bar{\rho}})^{\gamma_2-1} \right)^m \right]^{1/m} \\
  &=&  (C_{s,1}^0)^2 \left[   \left({\rho  \over \bar{\rho} }\right)^{(\gamma_1-1)m}  + \left({C_{s,2}^0\over C_{s,1}^0}\right)^{2m} 
\left({\rho  \over \bar{\rho}}\right)^{(\gamma_2-1)m}  \right]^{1/m},
\label{prescript}
\end{eqnarray}
where $m$ is a real number of order unity and where $C_s^0\equiv C_s^0(\gamma)$ is given by eqn.(\ref{csgam}). The exact dependence of the temperature upon density is 
not well known and may vary from place to place. Figure~\ref{temperature} shows the temperature
distribution as a function of the density.  
The two dash/dot-lines  correspond to the prescription of Jappsen et al. (2005).
In the following, we investigate  how the CMF depends on the 
critical density $\rho^{crit}$ and on the parameter $m$ which, as seen 
from fig.~\ref{temperature}, has some impact on the temperature variation. Low values of $m$ 
correspond to a smooth transition between the two regimes whereas larger values 
lead to a more abrupt transition.
As can be seen  from fig.~2 of Larson (1985), our knowledge of the temperature distribution is not very 
accurate and it is unclear  what would be the most appropriate choice for $m$.

Since at the critical density, the two contributions are equal, we have the relation:
\begin{eqnarray}
\left(C_{s,1}^0\right)^2 \left({\rho^{crit} \over \bar{\rho}}\right)^{\gamma_1-1} = \left(C_{s,2}^0 \right)^2 \left({\rho^{crit}  \over \bar{\rho} }\right)^{\gamma_2-1},
\end{eqnarray}
leading to
\begin{eqnarray}
\kcrit= \left({C_{s,2}^0 \over C_{s,1}^0}\right)^2 =    \left({\rho^{crit}  \over \bar{\rho} }\right)^{\gamma_1  -\gamma_2}.
\label{const_K}
\end{eqnarray}
The value of $\kcrit$ depends on the cloud's mean density $\nbar$ and on the critical density, ${\nbar}^{crit}$. 
Since molecular clouds have an average density of about $\nbar \simeq 100$ cm$^{-3}$, we find that
 the aforementioned value of ${\nbar}^{crit}$  leads to $\kcrit \simeq 0.06$.
Taking  a higher value, $\nbar \simeq 10^3$ cm$^{-3}$, which could be more representative of denser environments,
yields  $\kcrit \simeq 0.14$.

With these expressions, eqns.~(\ref{crit2}) and~(\ref{derive_M}) now become:
\begin{eqnarray}
\widetilde{M}_R^c &=&  \widetilde{R}  \left(  A  ^{1/m} + {\cal M}_*^2
\widetilde{R}^{2 \eta} \right), \\
{\rm with}\,\,\,A &=&  \left(  { \widetilde{M}_R^c \over \widetilde{R}^3 }   \right)^{(\gamma_1-1)m}  + 
( K_{\rm crit})^m \left( { \widetilde{M}_R^c \over \widetilde{R}^3 }   \right)^{(\gamma_2-1)m}, 
\nonumber
\label{multi_crit2}
\end{eqnarray}
and
\begin{eqnarray}
{d \widetilde{M}_R^c \over d \widetilde{R} } &=&  { B \over C},  \\  
\nonumber
{\rm with}\,\,\,B &=&  A^{\frac{1}{m}} +  (3-3 \gamma_1)  ({\widetilde{M}_R^c \over \widetilde{R}^3 } )^{m(\gamma_1-1)} A^{\frac{1}{m}-1}  +
(\kcrit)^m (3-3 \gamma_2)  ({\widetilde{M}_R^c \over \widetilde{R}^3 } )^{m(\gamma_2-1)} A^{\frac{1}{m}-1}  \\ 
&+&(2 \eta + 1) {\cal M}_*^2 \widetilde{R}^{2 \eta} , \nonumber \\
 C &=&  1 -  A ^{\frac{1}{m}-1} \times \nonumber 
\\ && \left( (\gamma_1-1) {(\widetilde{R}) \over \widetilde{M}_R^c }
({\widetilde{M}_R^c \over \widetilde{R}^3 } )^{m(\gamma_1-1)}
+ (\kcrit)^m (\gamma_2-1)  {(\widetilde{R}) \over \widetilde{M}_R^c } 
({\widetilde{M}_R^c \over \widetilde{R}^3 } )^{m(\gamma_2-1)} \right) ,
\nonumber
\label{derive_M2}
\end{eqnarray}
while eqns.~(\ref{mass_spec}),~(\ref{dens_crit}) and (\ref{derive_delt}) remain unchanged.

\begin{table}
\begin{tabular} {| {l} | {l} | {l} | {l} | {l} | {l} | {l} || {l} | {l} |{l} ||{l} | {l} | }
\hline
 $L$   & $d_0$ & $u_0$ & $\nbar$ & $\lambda_0$  &  $M_{cl}$ &  $V_{\rm rms}$   & ${\cal M}$ & 
$({\cal M}_*)^2$   & $L / \lambda_0$  & $E_k/E_p$ & $E_{th}/E_p$ \\
\hline
$ 0.5  $ & 3 & 1 & $ 4873  $ & 0.24 &  17                & 0.6  & 1.87 & 0.6    & 2.0  & 1.0 & 0.86 \\
\hline
$ 1    $ & 3 & 1 & $ 3000  $ & 0.33  & 88                & 0.8  & 2.39 & 0.7   & 2.98 & 0.76 & 0.4 \\
\hline
$ 2    $ & 3 & 1 & $ 1846  $ & 0.45  & 433               & 1.1  & 3.0  & 0.8   & 4.38 & 0.58  & 0.18 \\
\hline
$ 5    $ & 3 & 1 & $ 972   $ & 0.68 & 3570               & 1.6  & 4.2  & 1.0    & 7.3  & 0.4  & 0.067 \\
\hline
$ 0.5  $ & 10  & 1 & $ 16245 $  & 0.11 & 59              & 0.6  & 2.2 & 0.4   & 4.4  & 0.3 & 0.18 \\
\hline
$ 1  $ & 10 & 1 & $ 10000 $ & 0.15 & 293                 & 0.8  & 2.8 & 0.5    & 6.4  & 0.23 & 0.09 \\
\hline
$ 2 $ & 10 & 1 & $ 6155 $  & 0.21  & 1466                & 1.1  & 3.6 & 0.6  & 9.4   & 0.17  & 0.04 \\
\hline
$ 5 $ & 10 & 1 & $ 3241  $ & 0.32  & 11901               & 1.6  & 5.0 & 0.7    & 15.7 & 0.12 & 0.014 \\
\hline
$ 0.5 $ & 15 & 1 & $ 24367 $  & 0.088 & 89               & 0.6  & 2.3 & 0.4   & 5.7  &0.2 & 0.11 \\
\hline
$ 1  $ & 15 & 1  & 15000      & 0.16  & 440              & 0.8  & 3.0 & 0.45  & 12.2 & 0.15 & 0.05 \\
\hline
$ 2 $ & 15  & 1  & 9233       & 0.16 & 2169              & 1.1  & 3.8 & 0.5 & 12.0  & 0.11  & 0.02 \\
\hline
$ 5 $ & 15  & 1 &  4861       & 0.24 & 17851             & 1.6  & 5.3 & 0.6  & 20.3 & 0.08 & 0.0086 \\
\hline
$ 0.5  $ & 10 & 1.5 & $ 16245  $ & 0.11 &  59            & 0.9  & 3.3 & 1.0    & 4.4  & 0.68 & 0.18 \\
\hline
$ 1    $ & 10 & 1.5 & $ 10000  $ & 0.15  & 293           & 1.2  & 4.2 & 1.1   & 6.4 & 0.52 & 0.086 \\
\hline
$ 2    $ & 10 & 1.5 & $ 6155  $ & 0.21  & 1466           & 1.6  & 5.4  & 1.3   & 9.4 & 0.39  & 0.04 \\
\hline
$ 5    $ & 10 & 1.5 & $ 3241   $ & 0.32 & 11901          & 2.5  & 7.5  & 1.6    & 15.7  & 0.27  & 0.014 \\
\hline
$ 0.5  $ & 15  & 1.5 & $ 24367 $  & 0.11 & 89            & 0.9  & 3.5 & 0.9   & 5.7  & 0.45 & 0.11 \\
\hline
$ 1  $ & 15 & 1.5 & $ 15000 $ & 0.15 & 440               & 1.2  & 4.5 & 1.0    & 8.3  & 0.34 & 0.05 \\
\hline
$ 2 $ & 15 & 1.5 & $ 9233 $  & 0.21  & 2169              & 1.6  & 5.7 & 1.1  & 12.   & 0.26  & 0.02 \\
\hline
$ 5 $ & 15 & 1.5 & $ 4861  $ & 0.32  & 17851             & 2.5  & 7.9 & 1.4    & 20.3 & 0.18 & 0.009 \\
\hline
$ 0.5 $ & 15 & 2 & $ 24367 $  & 0.088 & 89               & 1.1  & 4.7 & 1.5   & 5.7  &0.8 & 0.11 \\
\hline
$ 1  $ & 15 & 2  & 15000      & 0.16  & 440              & 1.6  & 6.0 & 1.77  & 8.3 & 0.61 & 0.05 \\
\hline
$ 2 $ & 15  & 2  & 9233       & 0.16 & 2169              & 2.2  & 7.6 & 2 & 12.0  & 0.46  & 0.02 \\
\hline
$ 5 $ & 15  & 2 &  4861       & 0.24 & 17851             & 3.3  & 10.6 & 2.5  & 20.3 & 0.32 & 0.0086 \\
\hline
\end{tabular}
\caption{Parameters of clouds considered in fig.~\ref{comp_larson}. $L$, the size of the cloud
and  $\lambda_0$, the pseudo Jeans length, are in pc, $d_0$ is defined by eqn.~(\ref{def_dens}), 
$\nbar$, the average cloud density is in cm$^{-3}$, $M_{cl}$ the cloud mass is in solar mass whereas
 $V_{\rm rms}$, the root mean square 
velocity is in km s$^{-1}$. The parameters  ${\cal M}$, 
$({\cal M}_*)^2$ and  $L / \lambda_0$ which directly  enter the theory, are dimensionless. Finally
  $E_k/E_p$, $E_{th}/E_p$ represent the corresponding ratios of kinetic and thermal over gravitational energy.}
\label{param_dense}
\end{table}

\subsection{Peak position for a barotropic equation of state}
Finding an exact expression for the  position of the peak with a barotropic equation of state
is not possible. However, one can estimate its value in two limiting cases. 
First, if the contribution of the second term to the thermal pressure (eqn.~(\ref{prescript})) is negligible, 
i.e. $\kcrit \ll 1$, the eos is nearly polytropic and the peak position is given by 
eqn.~(\ref{minimum}). Second, if for the densities corresponding  to  the mass of the peak position, $\simeq M_{\rm peak}/R^3$,
the second contribution of  eqn.~(\ref{prescript}) is dominant, 
then in the thermally dominated regime, we have with eqn.~(\ref{multi_crit2}):
\begin{eqnarray}
\widetilde{M}_R^c \simeq \left( K_{\rm crit} \right)^{1 \over 2 - \gamma_2}
\widetilde{R}^{4 - 3 \gamma_2 \over 2 - \gamma_2}.
\label{M_R_therm}
\end{eqnarray}
and the peak position occurs at:
\begin{eqnarray}
\label{minimum_baro}
\mp &=& \kcrit ^{ 6 - 3 \gamma_2 \over 2 (2-\gamma_2)} 
\exp \left( { (4-3 \gamma_2) (-9+6 \gamma_2) \sigma^2  \over 4} \right) \\
&=& \left( {\bar{\rho} \over \rho^{crit}} \right) ^{ (\gamma_2 - \gamma_1) (6 - 3 \gamma_2) \over 2 (2-\gamma_2)}  (1 + b^2 {\cal M}^2) ^{-(4-3 \gamma_2) (9-6 \gamma_2)  \over 4}.
\nonumber
\end{eqnarray}
In any case, the peak position should be close to the maximum of the two values given
by eqns.~(\ref{minimum}) and (\ref{minimum_baro}), respectively:
\begin{eqnarray}
\label{minimum_approx}
\widetilde{M}_{\rm peak} \simeq \max \left(
 \left( {\bar{\rho} \over \rho^{crit}} \right) ^{ (\gamma_2 - \gamma_1) (6 - 3 \gamma_2) \over 2 (2-\gamma_2)}  
(1 + b^2 {\cal M}^2) ^{-(4-3 \gamma_2) (9-6 \gamma_2)  \over 4}, (1 + b^2 {\cal M}^2) ^{-(4-3 \gamma_1) (9-6 \gamma_1)  \over 4} \right)
\end{eqnarray}

Figure~\ref{mass_peak} shows the peak position as a function of $\exp(\sigma^2)=1+b^2 {\cal M}^2$. 
The solid lines display the peak position (eqn.~(\ref{minimum_baro})) for $\gamma_2=1.1$ 
and for various values of $\rho^{crit}/{\bar \rho}=$1, 10, $10^2$ and 
$10^3$. The dashed line corresponds to the 
peak position  for $\gamma_1=0.7$ (eqn.~\ref{minimum}) whereas the dotted line 
is for $\gamma_1=1$ (isothermal gas). For realistic values of  $\exp(\sigma^2) \simeq 5-50$, the peak position 
is almost always given by the solid lines since they lead to values larger than the 
values displayed by the dashed line. Therefore, the peak position is essentially
 determined by the {\it nearly isothermal 
   regime} (effective polytropic exponent $\gamma_2 \simeq 1.1$ in this case). This stems from the fact that 
as shown in fig.~\ref{fig_gamma}, the peak position changes very rapidly with the value of $\gamma$. 

The comparison between the dotted and solid lines reveals that, for realistic values
of the ratio $\rho^{crit}/\bar{\rho} \simeq 10^2-10^3$, the peak position occurs at much smaller masses in the 
barotropic case considered here than in the isothermal case. This indicates that the cloud's mean density has to be
much smaller in the barotropic case than in the isothermal case for the peak of the CMF to 
occur at the same mass. 

It is already interesting at this stage (detailed comparisons are performed
in \S~\ref{comp_simu}), to compare these analytical results with the 
numerical results obtained by Jappsen et al. (2005), displayed in their figure~5, 
which portrays the IMF obtained for various values of ${\nbar}^{\rm crit} / \nbar \simeq$ 0.5, 5, 50 and
 500,  corresponding to
$\nbar \simeq 8.4 \times 10^4$ cm$^{-3}$ and ${\nbar}^{\rm crit}= 4.3 \times 10^4, \;
4.3 \times 10^5, \; 4.3 \times 10^6$ and $4.3 \times 10^7$ cm$^{-3}$, respectively.
We estimate the peak  of the IMF for these four values of 
${\nbar}^{\rm crit} / \nbar$ to be located at respectively $M_{peak}/M_\odot\simeq 2.5$, 
0.5, 0.16 and 0.06, in solar mass units.
These numbers can be compared with eqn.~(\ref{minimum_baro}) which, for $\gamma_1=0.7$ and 
$\gamma_2=1.1$, predicts that $\mp \propto (\rho_{\rm crit} / {\bar \rho}) ^{0.6} $.
Computing the corresponding exponent from the above quoted values, we find typically 
$\mp \propto (\rho_{\rm crit} / {\bar \rho}) ^{0.55} $ in the simulations
which, given the large uncertainties inherent to such difficult numerical calculations, appears to be 
fully compatible with the value of 0.6 predicted by eqn.~(\ref{minimum_baro}).

\subsection{Results for a barotropic equation of state}
Figure~\ref{fig_gamma_var} shows the core mass function  for  ${\cal M}= 6$ and ${\cal M}_*^2= 2$,
for various values of the constant $\rho^{crit}/\bar{\rho}$, namely 10, $10^2$ and $10^3$. 
The solid lines display the case $m=1$ in eqn.~(\ref{prescript}), the dashed lines correspond 
 to $m=3$ whereas the dotted line corresponds to the isothermal case. 
As expected, the number of high-mass stars
does not depend on the gas thermodynamics whereas the number of low-mass objects 
increases with increasing values of $\rho^{crit}/\bar{\rho}$. For $\rho^{crit}/\bar{\rho} \simeq 10$, 
$\gamma_1=0.7$ and $\gamma_2=1.1$, the CMF is close to the isothermal case. 
Figure~\ref{fig_gamma_var} also shows the non negligible dependence of the CMF upon the parameter
$m$, which describes the - ill defined - transition between the two cooling regimes.

\begin{figure}[p]
\center{\includegraphics[angle=0,width=5in]{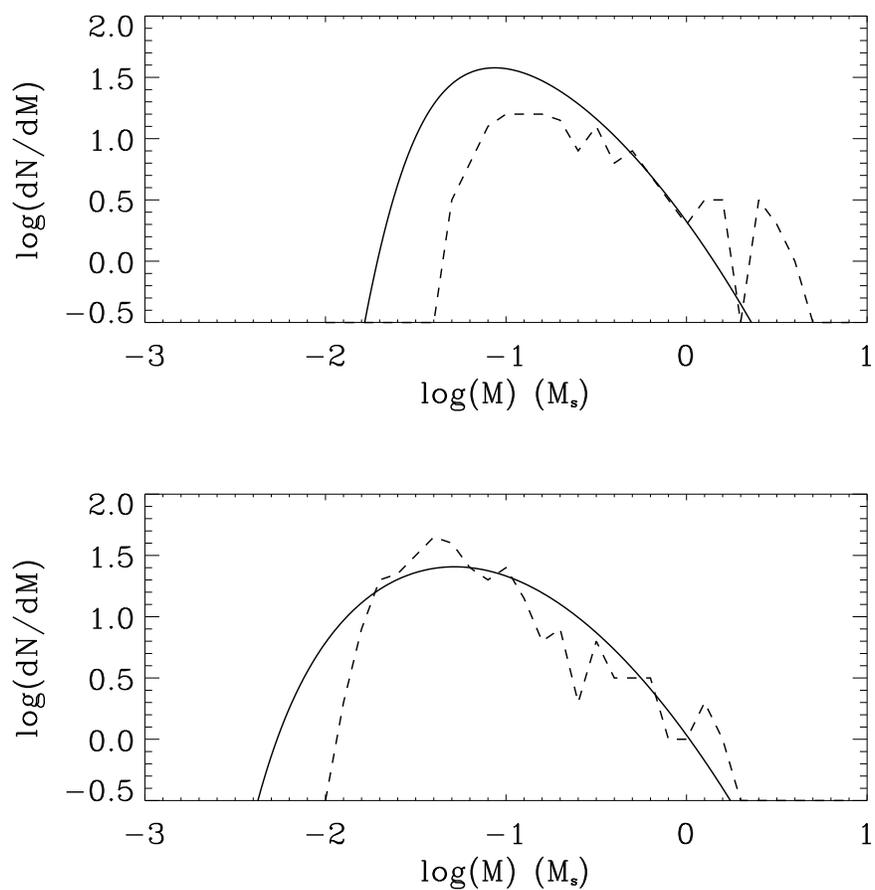}} 
\caption{Comparison between our analytical IMF and the
 results of Jappsen et al. (2005) obtained using SPH 
simulations and sink particles, for the same characteristic parameters (see text). Top panel: $\ncrit = 4.3 \times 10^6$ cm$^{-3}$; bottom panel: $\ncrit = 4.3 \times 10^7$ cm$^{-3}$.}
\label{comp_jappsen}
\end{figure}

\begin{figure}[p]
\center{\includegraphics[angle=0,width=8in]{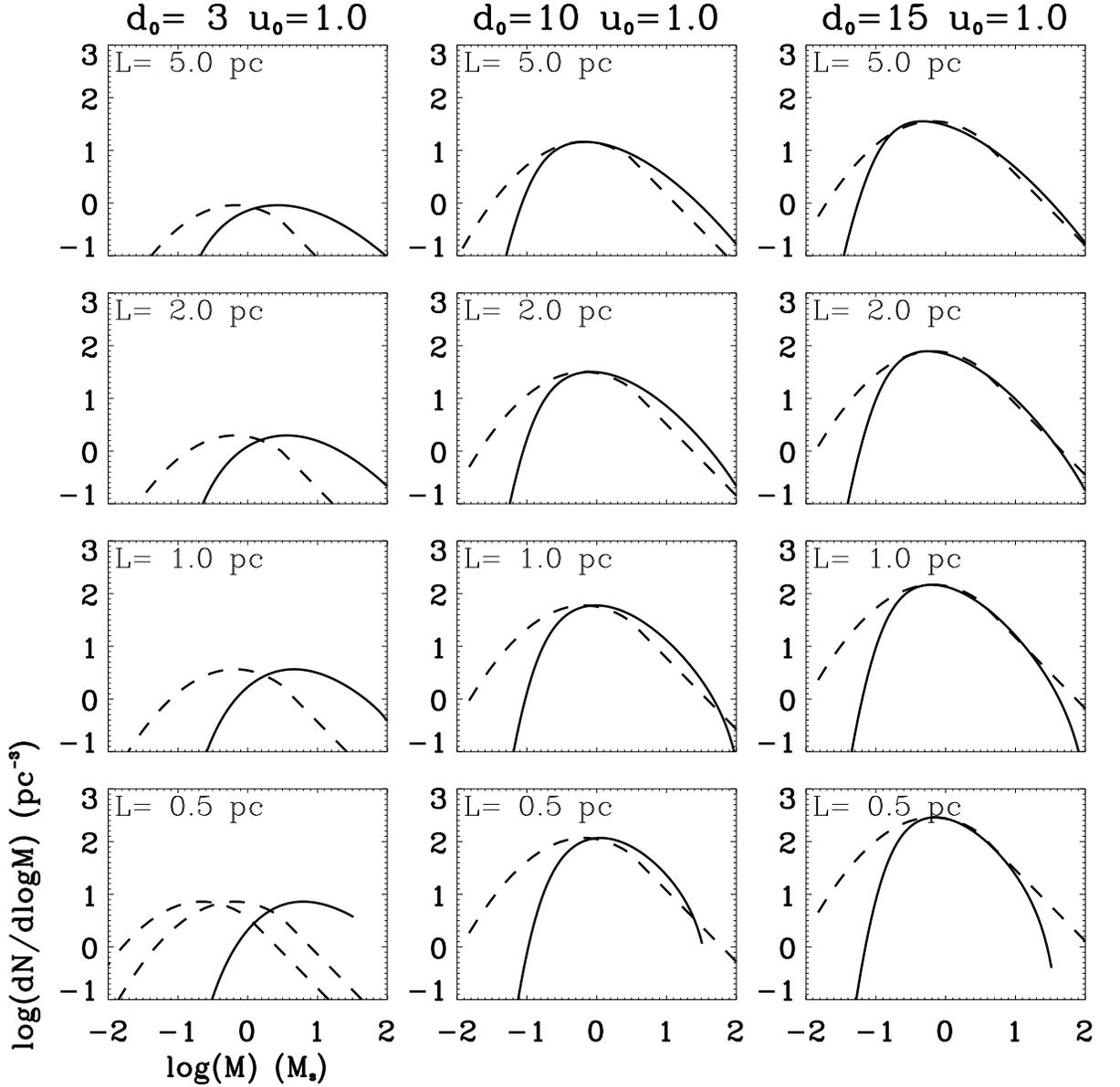}} 
\caption{Comparison between the Chabrier's IMF (dashed line) and the 
CMF predicted in the present theory for clouds following Larson-type relations (see text), for $u_0=1$ km/s (see eqn.(\ref{def_dens})). The dash-curve which peaks at about $1 \, M_s$
is the Chabrier's IMF shifted upward by a factor 3 in mass to account for the shift between the CMF and the IMF. In the bottom-left panel the second 
dashed curve shows the original (non shifted) Chabrier's IMF.}
\label{comp_larson}
\end{figure}

\begin{figure}[p]
\center{\includegraphics[angle=0,width=8in]{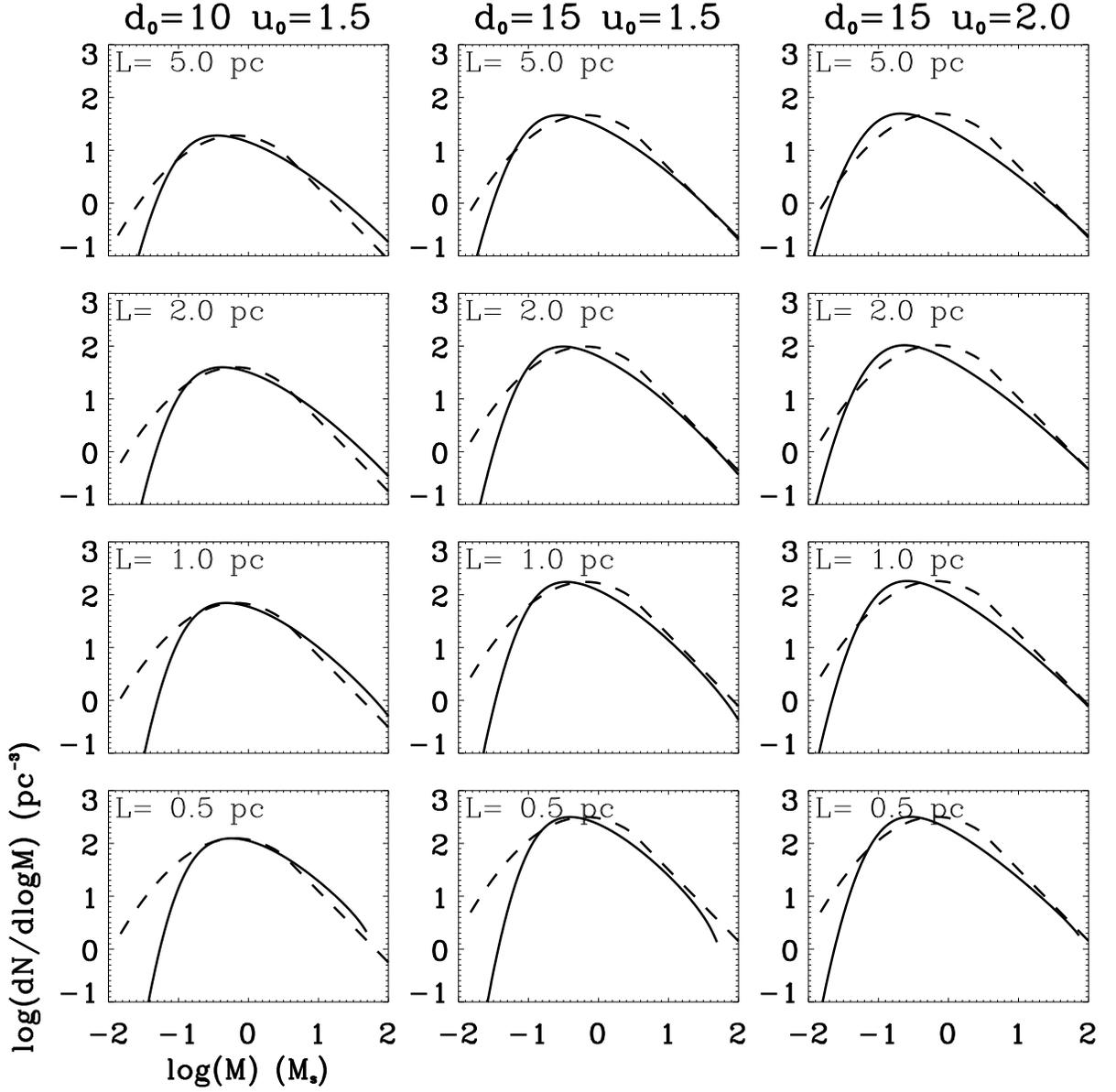}} 
\caption{Comparison between the exact CMF/IMF and the approximated one 
(given by eqns.~(\ref{crit_approx_deriv})-(\ref{crit_approx_norm}). The solid line and dashed lines 
show respectively the exact calculation for $m=1$ and $m=3$ while the dotted line shows the approximated 
CMF/IMF.}
\label{compar}
\end{figure}


\begin{figure}[p]
\center{\includegraphics[angle=0,width=6in]{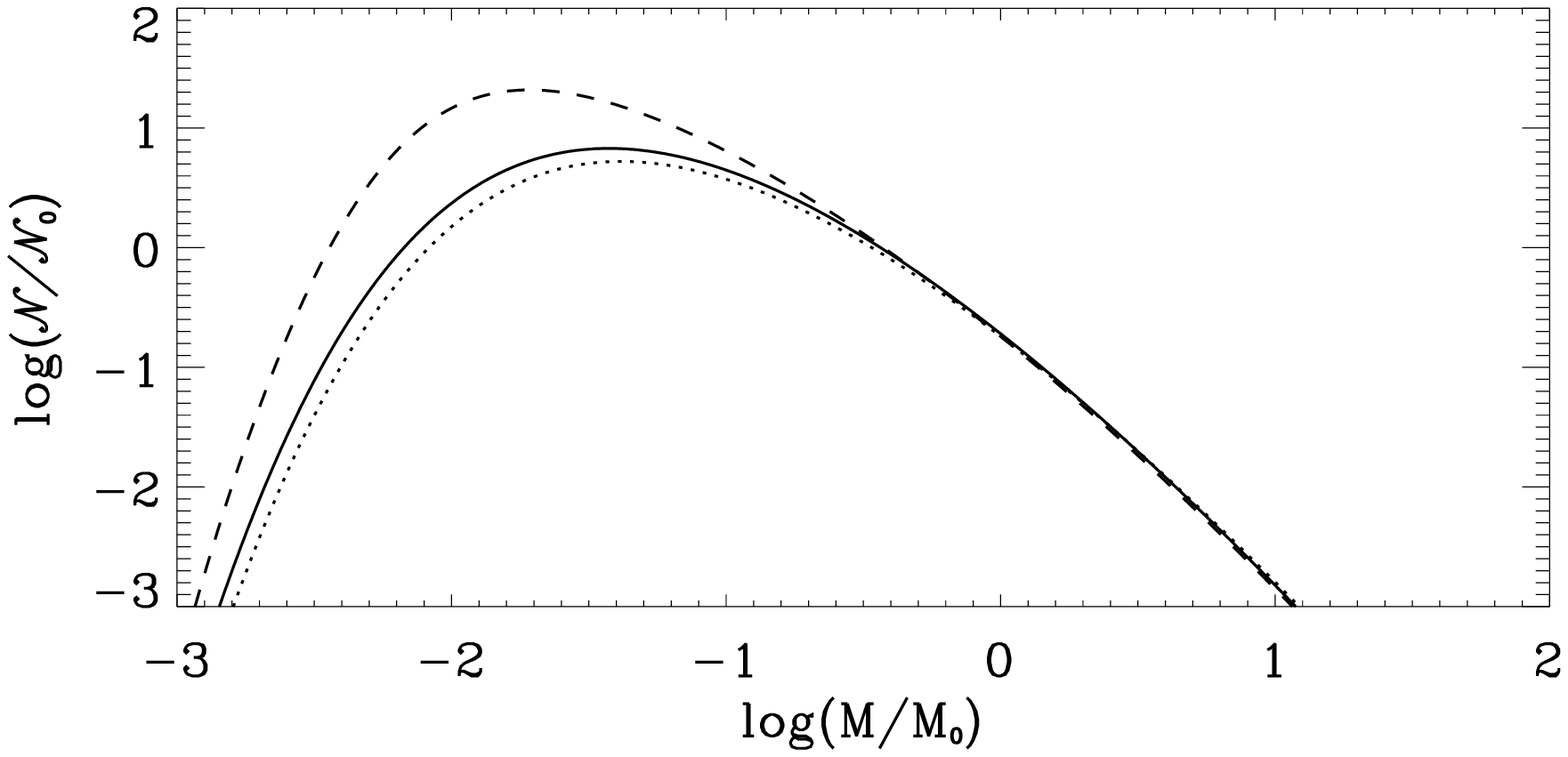}} 
\caption{Same as fig.~\ref{comp_larson} for $d_0=10$ and 15 and for various values of 
 $L$ and $u_0$ (=1.5 and 2).}
\label{comp_larson2}
\end{figure}

\section{Comparison with numerical simulations and observations}
In this section, we carry out detailed comparisons between our theory and the IMF determined both by the observations and 
by the numerical simulations performed by Jappsen et al. (2005) who determine the mass spectrum of collapsed objects in their 
simulations. 

\subsection{Comparison with numerical simulations}
\label{comp_simu}
In order to compare the simulations of Jappsen et al. (2005) with the present theory, we need to 
estimate the values of the parameters ${\cal M}$, ${\cal M}_*$ and $\kcrit$ in the simulations.
Since the mean density in the simulations is $8.4 \times 10^4$ cm$^{-3}$ while 
$C_s=0.2$ km $^{-1}$, the pseudo Jeans length 
(recalling that the pseudo Jeans length differs from
the standard definition by a constant factor and assuming $a_J= \pi^{5/2} /6$) is about 
$\lambda^0_J = (\sqrt{\pi} /2)  C_s/\sqrt{G \bar{\rho}} \simeq 0.042$ pc.
The box size is 0.29 pc, giving $L/\lambda^0_J=6.9$. The Mach number in the simulation is 3.2 which 
gives ${\cal M}_* = {1 \over \sqrt{3} } {\cal M} (\lambda_J^0/L)^\eta \simeq 0.8$. 
 Finally, Jappsen et al. (2005) explore various values of $\ncrit$. For
 $\ncrit= 4.3 \times 10^6$ cm$^{-3}$ (fourth row of fig.~5 of Jappsen et al. 2005), this yields
 $\rho_{\rm crit} / {\bar \rho} \simeq 50$ while 
for $\ncrit = 4.3 \times 10^7$ cm$^{-3}$ (fifth row of their fig.~5), this yields
 $\rho_{\rm crit} / {\bar \rho} \simeq 500$. Finally, we adopt the value $m=3$ which leads to 
a temperature distribution very close to the one adopted by Jappsen et al. (2005) and a
value of  $b=0.5$ typical of the type of  forcing  used in these simulations. 
We stress that our goal in this section is not to determine 
the most accurate IMF but rather to test the predictions of the analytical model with the direct 
numerical simulations.

Dotted lines of fig.~\ref{comp_jappsen} show the numerical results of Jappsen et al. (2005)
for the cases $\ncrit = 4.3 \times 10^6$ cm$^{-3}$ (top panel) and $\ncrit = 4.3 \times 10^7$ cm$^{-3}$
(bottom panel) while solid lines display the prediction of the present theory with the aforementioned
 parameters. 
The CMF predicted by the theory has been arbitrarily normalized to the numerical one  near 1 $\msol$.
No adjustment has been made along the x-axis, implying that 
the peak position results from the dimensionless numbers (mainly Mach number) and the Jeans 
mass, including the coefficient $a_J$. 
The agreement between the analytical theory and the numerical calculations is 
encouraging and suggests that the theory is indeed grasping the essential of the physics
occurring in the simulations. In particular, it can be seen that in both cases, the peaks are 
approximately at the same place.
It seems, however, that the analytical 
model is producing more brown dwarfs than the simulations. A possible explanation
is that the value of $a_J$ should be modified. Indeed, $a_J=\pi^{5/2} /6$ is obtained 
by computing the mass enclosed within a sphere whose diameter is the Jeans mass. Although reasonable, 
 such a definition is not very accurate and the comparisons
suggest that we should take 
a value for $a_J$ about twice larger. Other, not exclusive,
 possibilities are that either the theory or the numerical simulations are
 not accurate enough in this regime.

 While the slopes at high masses are   similar,  it is maybe 
the case that the analytical curve never ceases to have a curved shape while the distribution inferred from the 
numerical simulations display
a clearer powerlaw behaviour, particularly for the second panel. We note however that the numerical simulations 
appear to be a bit noisy, preventing to reach a definite conclusion at this stage
and this behaviour has to be confirmed by further simulations.
One possibility would be that 
a more accurate  density PDF should be used or that accretion onto the sink particles modify the shape of the 
distribution.

\subsection{Comparison with observations}
\subsubsection{Cloud and model parameters}
 Using the barotropic eos specified in \S~\ref{barotrop}, our fiducial model is defined as,
$C_s \simeq 0.26$ km s$^{-1}$ for a density of $n = 10^4$ cm$^{-3}$ 
(implying a temperature of about 20 K at this density),  and 
$m=2$ (see eqn.~(\ref{prescript})), 
 $\ncrit =4 \times 10^5$ cm$^{-3}$, $\gamma_1=0.7$, $\gamma_2=1.07$, and setting $b=0.75$ and $\eta=0.45$.
The theory  entails three main parameters, $\bar{\rho}$, ${\cal M}$, and ${\cal M}_*$, 
respectively the average cloud density, the Mach number at the cloud size and the 
Mach number at the mean Jeans length.
Note that $\bar{\rho}$ enters the constant $\kcrit$ stated by eqn.~(\ref{const_K}) and also 
the normalization, in the definition of $M_J^0$. Finally, as mentioned previously, we
choose $a_J=\pi^{5/2} /6$.

Three parameters still represent a significant range of possibilities which are not all worth 
exploring, so we will restrict ourselves to a smaller subset. Since it is well 
established that interstellar clouds seen in the CO lines follow Larson-type  relations, 
it seems natural to consider only  clouds whose characteristics obey
such relations. As recalled in \S~\ref{theory}, the Larson relations (Larson 1981) correspond to 
$V_{\rm rms} \simeq 0.8 \, {\rm km \, s}^{-1} \; ({\rm L/1 pc})^\eta$, 
$\nbar \simeq 3 \times 10^3 \, {\rm cm}^{-3} ( {\rm L/1 pc})^{-1}$, where $\eta \simeq 0.4-0.5$. 
Therefore, once the size of the cloud is specified, all the relevant parameters can be
 inferred. Such a size-velocity dispersion scaling relationship has been confirmed by several authors
(e.g. Heithausen et al. 1998, Falgarone et al. 2004) in various regions, although a value $-0.7$ in the size-density relation, instead of $-1$, seems to provide a better fit to the data.
The mean density at a given size, however, 
appears to undergo significant variations. For example, the data compilation presented in 
Falgarone et al. (2004) seems to indicate that the typical density at a given size can vary 
from one region to another over at least one order of magnitude.
This leads us to consider clouds of mean density  defined as:
\begin{eqnarray}
\nbar = (d_0 \times 10^3 \; {\rm cm}^{-3})  \left( { L \over 1 {\rm pc}} \right)^{-0.7},\,\,
V_{\rm rms} = (u_0 \times 0.8 \; {\rm km \; s}^{-1}) \left( { L \over 1 {\rm pc}} \right)^{\eta}.
\label{def_dens}
\end{eqnarray}
Note that theoretically, the existence of a density-size relation  has been questioned by Kegel (1989), V\'azquez-Semadeni et al. (1997)
and  Ballesteros-Paredes \& MacLow (2002). More recent works have nevertheless obtained clear density-size relation
(Hennebelle et al. 2007, Kritsuk et al. 2007). In any case, this relation is not at all fundamental in the present context 
and simply used as a simple prescription to determine the non dimensional parameters which control our theory.

In the following, we will consider four different cloud sizes, namely
0.5, 1, 2 and 5 pc  representative of the  size of a clump in  a  GMC. 
We restrict ourselves to 3 values of $d_0$, namely 3, 10 and 15,
which correspond to clouds respectively as dense as and  denser than the value originally 
quoted by Larson. 
 Since it is interesting to explore also the dependence of our results upon the velocity dispersion, 
we also consider the case of various velocity dispersions, namely $u_0=1, 1.5$ and 2. 
Indeed, the recent study of Heyer et al. (2008) suggests that the velocity dispersion 
is correlated with the column density of the cloud $\Sigma$ as $u_0 \propto \Sigma^{1/2}$.  
Table~\ref{param_dense} lists the various physical parameters of our cloud models, the 
 dimensionless parameters which enter in the theory, as well as  the ratios $E_{\rm th}/E_p $ and $E_k/ E_p$ where
 $E_{\rm th}=(3/2) M C_s^2$ is the thermal energy, $E_k = (1/2) M V_ {\rm rms}^2 $ the 
kinetic energy and  $E_p= (3/5) G M^2/(L/2)$ is the gravitational energy.


\subsubsection{Model versus observations}
Figure~\ref{comp_larson} portrays the comparison between the  models (solid lines) described 
in the previous section and 
the Chabrier's {\it system} IMF (dashed lines, Chabrier 2005). Indeed, we recall that the present theory predicts the distribution of
prestellar cores, as observed in dust-continuum surveys, and does not include further subfragmentation of these cores into individual objects (see discussion in \S7.1.2 of paper I). In the same vein, the resolution of present observational surveys does not allow to resolve individual objects in the observed bound cores.
Since there is observational evidence that the CMF 
appears to be shifted with respect to the IMF by a constant factor of about 3 in mass (Motte et al. 1998, 
Alv\'es et al. 2007),
the Chabrier's IMF is shifted in mass by the same factor on the figures. 
The Chabrier's IMF is normalized to the observed stellar density near 1 $\msol$ in the solar neighborhood but the normalization in our calculations is arbitrary.
 Therefore, in order to compare the observational and theoretical curves, they have been 
adjusted along the y-axis at the same peak value. We emphasize
that no adjustment has been made along the x-axis (mass). That is, for a given analytic distribution $\cal {N} (\widetilde{M})$, we plot 
$d N / d \log M = \ln(10) M \cal {N} ( M)$, where $M = M_J^ {0 ^{\prime}} \widetilde{M}$.

For $d_0=3$, which corresponds to the value originally quoted by Larson,
 the agreement between the theoretical CMF and the Chabrier's IMF is poor.
The two main important disagreements are the peak value, which is  larger for the 
model than for the shifted Chabrier's IMF, and the slope at large masses which is 
not stiff enough for the model. While the first discrepancy is simply due to the Jeans mass
being too large,   the second  discrepancy can be understood as follows. 
Equation~(40) of paper I shows that at large masses, the value of $x$ is slightly 
more complex than indicated by eqn.~(42) of paper I, and is exactly given by: 
\begin{eqnarray}
x={n + 1 \over 2 n - 4} - {6 (1-\eta) \over 2 \eta +1} {\ln { \cal M}_* \over \sigma ^2}.
\label{x_salp}
\end{eqnarray}
Although the value of the second term is small when ${\cal M}_* \simeq 1$ and 
${\cal M} \simeq 6-10$, it can lead to non negligible corrections when ${\cal M}_* $
takes larger values  and ${\cal M}$ smaller values, as is the case here for some of the cloud conditions listed in 
table~1, for which the second 
term can amount to about 0.2, leading to $x \simeq 1.1$.

For $d_0=10$ and $u_0=1$,
a better although imperfect agreement between the observations and the prediction of the present
 theory is obtained. We see that the peak position, even though closer to the peak of the
shifted Chabrier's IMF, does not exactly coincide with it. 
The slope at large masses is very close to the Salpeter value,
$x \simeq 1.35$, but too many massive stars are produced. There is also an obvious 
disagreement in the low-mass regime where the model predicts significantly less 
brown dwarfs and low-mass stars. The agreement is  improved if we  consider a 
higher velocity dispersion, 
$u_0=1.5$, as revealed by the first column of  fig.~\ref{comp_larson2}. The peak is now approximately 
at the right place whereas the number of high mass stars is in better agreement, although still slightly 
larger than the observed values. As noted in \S7.1.4 of paper I, 
the peak of the IMF does not change significantly when changing the size of the clumps, assuming they obey
Larson-type relations. This is due to a cancellation
between the  increasing Jeans mass and the increasing Mach number as the cloud size increases.

In the case $d_0=15$, a better agreement between the theory and the shifted Chabrier's IMF 
is obtained both for $u_0=1$ (fig.~\ref{comp_larson}) and $u_0=1.5$ (fig.~\ref{comp_larson2}). 
The peak and the high mass part of the distribution are very well reproduced 
by the models except perhaps for $L=0.5$ pc and $u_0=1$, for which the number of 
high mass stars may drop too stiffly (because of the too low global Mach number $\cal M$). The lack of brown dwarfs is less severe than for 
the case $d_0=10$. For the cases $L=2$ and 5 pc and $u_0=1.5$, in particular, 
the remaining disagreement is occurring at very low masses
where the observations are not well constrained. 
Finally, in order to investigate the influence of larger values of $u_0$, we also present 
(third column of fig.~\ref{comp_larson2}) the case $d_0=15$ and $u_0=2$. 
In that case, the peak occurs at too small masses while the slope in the 
high mass part is  slightly too shallow.

\subsection{Discussion}

According to the predictions of our theory, we conclude that typical conditions for 
star formation should differ from those specified by the Larson's relations. In particular, 
the clouds characteristic of star forming regions need to be denser than predicted by these relations. However, two points must be stressed.
First of all, while comparing our results with the shifted Chabrier's IMF (based on the observed shift between the CMF and the IMF),
we choose a constant value of 3 for the shifting factor. Although such a value has been inferred from observations in the Pipe nebula (Alfv\`es et al. 2007), it may in fact
depend on the cloud and exhibit some variation. For instance, Motte \& Andr\'e (1998) find a value closer to 2 in the rho Ophiuci cloud, while J\o rgensen et al. (2007) find a $\sim 10$-15\% core-to-star efficiency in Perseus, implying a $\sim 7$ to 10 shifting factor between the CMF and the IMF, although such numbers should be taken with due caution. Second of all, our theory
  ignores the magnetic support, which would add up to thermal and turbulent support. 
As discussed in \S4.4 of paper I, magnetic support can enter our theory by simply rescaling the turbulent support, and thus the effective Jeans length/mass, according to the observed correlation between the magnetic intensity and the density and rms velocity dispersion in molecular clouds, affecting also to some extend the Mach number and thus the width of the lognormal distribution. However, such a treatment is certainly a crude approximation of the true impact of magnetic field. 
At any rate, taking magnetic field into account would increase the support, implying denser clumps. 
 All these points suggest that the clumps in which star formation is taking place should be at least as dense as, probably
denser than the $d_0=15$ case displayed in the previous section. 
This suggests that star formation should occur preferentially in dense gas regions 
 where the gas density is typically 3-5 times larger than the canonical values originally inferred by Larson for the CO clumps.
Note that this must be the case at all scales which, in principle, arises in a turbulent flow.
 Indeed, polytropic flows are scale free and  varying the mean density implies varying the density at all scales.
For a clump of 1 pc, the density for which we get reasonable agreement is of the order of 
$\gtrsim 1.5\times10^4\cc$ according to our analysis.
 This may suggest that a triggering mechanism might be necessary to lead to such high mean densities.
This is consistent with a picture of star formation happening in relatively dense clumps, as suggested by many 
observations of star forming regions. In
Perseus, Serpens, Ophiucus, for instance,  cores are found preferentially at high cloud column densities (Av $>$ 6), corresponding to mean densities
${\bar n}>2\times 10^4\cc$ (Enoch et al. 2008). These conclusions are also consistent with the low star formation 
efficiency observed in the Galaxy since they imply that only a tiny fraction of the gas is actually 
forming stars efficiently.

There is a possible trend for the models to form too few low-mass brown dwarfs. Although quantifying the exact amount of this discrepancy requires
more sound determinations
of the IMF in the low-mass brown dwarf regime, its
origin, if true, is unclear. One possibility is that the thermodynamics of the 
gas is different from what has been assumed here. In particular, if the critical density, 
$\rho^{\rm crit}$, 
or the  effective adiabatic exponent, $\gamma_2$, are respectively higher and/or smaller than the value used in the present study, 
the number of small
mass objects will increase. Since the thermodynamic properties of the high density gas are not well known 
this possibility can not be excluded. Another suggestion is that gravity, which appears in our theory only under the form of a Jeans
criterion to define gravitationally bound objects, might lead to more fragmentation in the very-low-mass regime. Similarly, a lognormal distribution for
the turbulence-driven density fluctuations and a variance given by eqn.(\ref{sigma_val}), in particular in the non-isothermal and/or magnetized case, might be an over-simplification. As mentioned earlier, simulations including compressive forcing, for instance, clearly show a departure from such a distribution, with a PDF extending further down into the small-scale regime (Federrath et al. 2008).
In the same way, the density  PDF that we consider is valid in the isothermal case but should clearly 
be modified when non isothermal eos are considered. Let us recall that Passot \& V\'azquez-Semadeni (1998), find that 
the  PDF is not a lognormal and present a powerlaw tail at high densities when $\gamma < 1$. Taking this effect into account
 would certainly increase the number of brown dwarfs. Finally, it sounds likely that gravity will also lead
to a similar effect with the high density tail of the PDF becoming less stiff and more extended (Klessen 2000, 
 V\'azquez-Semadeni et al. 2008, Hennebelle et al. 2008). 
Indeed, the role of gravitational focussing in the production of brown dwarfs has been explicitly investigated by  Bonnell et al. (2008).

The global success of the present theory reinforces the suggestion that gravoturbulent fragmentation seems to be the dominant global physical mechanism responsible
for star and brown dwarf formation. This is supported, on the theoretical side,
by the general fairly good quantitative agreement, over basically the entire mass spectrum, between the observationally determined and the theoretical CMF/IMF, as illustrated
in the previous sections, and on the observational side by various explorations of young brown dwarf properties (Luhman et al. 2007, Duch\^ene et al. 2009) and by the recent statistical determination of the brown-dwarfs to stars ratio in several young clusters (Andersen et al. 2008). This latter study
shows that a Chabrier IMF, continuously extending from the stellar to the brown dwarf regime, is fully consistent with the observations.
It is not excluded, however, that the low mass end of the IMF might be affected by secondary mechanisms such as, for instance, dynamical ejections and radiative feedback (Bate 2009) or massive disk fragmentation (Stamatellos \& Whitworth 2009). Concerning the second mechanism, it should be stressed, however, that these calculations do not
include magnetic field, which has been shown to severely inhibit disk fragmentation (Machida et al. 2005, 2008, Hennebelle \& Teyssier 2008, Price \& Bate 2008) or even to prevent disk formation 
by magnetic braking (Galli et al. 2006, Price \& Bate 2007, Hennebelle \& Fromang 2008, Mellon \& Li 2008a,b). Interestingly, Machida et al. (2008)
 find that the formation of low mass objects is possible during the second collapse after 
the gas and the magnetic field decouple. 


\section{Conclusion}
In this paper, we have further investigated  the analytical theory we have recently developed for turbulence-driven star formation (Hennebelle \& Chabrier 2008). We have explored and quantified the influence 
of various properties of the flow, such as the turbulent forcing or the fluctuations of the rms velocity field, on the mass spectrum. Moreover, we have extended our previous calculations, based on an isothermal equation of state for the gas, to a more general 
polytropic or barotropic case. We show that the low mass
end of the CMF/IMF is very sensitive to the value of $\gamma$, the effective adiabatic 
index,  and to the temperature-density dependence within the cloud (see Fig. 4 and 6). Departure from isothermality with smaller values of $\gamma$ ($\gamma<1$) leads to a more compressible flow and cooler overdense structures, increasing the number of small-mass collapsing cores, precursors of brown dwarfs. Extensive comparisons with the non-isothermal numerical simulations performed by Jappsen et al. (2005) show the
good agreement between the IMF obtained in the simulations and the predictions of the theory.

Taking into account the thermal properties of the flow and a more complete description of turbulence 
forcing  improves the {\it quantitative} agreement between the theoretical CMF/IMF and the observationally-derived
 IMF, for more realistic Mach number values,  typical of star forming regions. A proper combination of these two 
properties, through the $\gamma$ and $b$ coefficients, leads to an excellent agreement between theory and observation.
Unfortunately, the lack of knowledge of both the precise thermodynamic properties of the flow at high density and of the PDF of non-isothermal
turbulence prevent an accurate determination of these two parameters under star-forming cloud conditions. It is not excluded, in particular, that the theory shows a trend for a deficiency of brown dwarfs. As mentioned above and in paper I, however, observational determinations of the brown dwarf IMF in the low-mass
regime are still very uncertain and more accurate determinations are required before allowing robust conclusions.
Various comparisons between our models and the observed IMF, for different cloud conditions, suggest that star formation 
should dominantly occur in clouds typically 5 times denser than the CO clumps characterized by Larson (1981), a point which seems to be supported by observations. 
This raises the question of the universality of the IMF since choosing different cloud parameters would lead to 
different CMF/IMF, in contradiction with the observational evidence. The most plausible answer to this apparent puzzle
lies in the universal nature of the turbulence spectrum, which tends to produce clouds with similar characteristics (mass versus size relation, velocity dispersion...), with
efficient star formation, i.e. formation of gravitationally bound cores, only or preferentially occurring in the clouds exceeding some mean density value.
Note that in the present theory, the characteristic mass of the IMF/CMF is set up by the cloud's {\it general conditions}. Meaning it
depends both on the average density and temperature (thus mean thermal Jeans mass) of the cloud but also on its characteristic (large-scale) Mach number, illustrating the combined effect of gravity and turbulence to set up the IMF. However, as demonstrated in \S7.1.4 of paper I and discussed in \S5.2 of the present paper, the
similar but opposite dependence of the Jeans mass and Mach number upon the cloud's size/mass, for clouds obeying Larson's type relations, leads
to an IMF peak position and characteristic mass, which arise in the nearly isothermal regime (see \S 5.2), nearly independent of the cloud's conditions. This is in contrast to the suggestions of Elmegreen et al. (2008) and Bate (2009). While the former authors invoke the weak environmental dependence of the ocurence of gas-dust coupling to
explain the universality of the characteristic Jeans mass, the second author argues for radiative feedback of prestellar cores to yield an effective Jeans
mass nearly independent of the cloud's density. Interestingly, these two approaches yield opposite density-dependences of the Jeans mass.
More detailed comparisons between dedicated simulations and the present theory are necessary to
explore this "universality" issue. 

An already interesting success of the present theory, testable with numerical simulations, is that it provides a consistent 
explanation to why gravitational clumps obtained by taking into account purely thermal support lead to too steep an IMF at 
large masses (Padoan et al. 2007), while simulations  which use  sink particles and include by definition the turbulent
support with a Kolmogorov-like power spectrum index obtain the proper Salpeter slope (e.g. Klessen 2001, Bate \& Bonnell 2005). 
Furthermore, as explored in \S6.3 of paper
I, this theory shows that turbulence globally {\it decreases} star formation efficiency, explaining at least partly the low 
efficiency of star formation observed in the Galaxy.

The overall good {\it quantitative} agreement between the present results and the CMF/IMF derived both from numerical simulations (including compressible-mode forcing
and non-isothermal gas eos) and observations over basically the entire mass spectrum, down to about 0.01 $M_\odot$, although by no means an
irrefutable proof, reinforces the
validity of the present theory. This latter suggests that gravo-turbulent fragmentation,
including the two {\it opposite} roles of turbulence, turbulence-driven density enhancements at the cloud's scale on one hand and turbulent support at the core's scale on the other hand, 
is a promising {\it dominant} physical mechanism responsible for star {\it and} brown dwarf formation. Alternative mechanisms very
likely contribute only marginally to this dominant process at the stage of prestellar core formation. For instance, observations (Andr\'e et al. 2007, Evans et al. 2008) suggest that collision timescales between cores are substantially longer than their lifetimes, so that
collisions and thus competitive accretion between dense cores are unlikely to be dominant. This in turn suggests that the prestellar dense cores are likely to evolve individually into stellar systems. This property lies at the very heart of the present theory. As mentioned earlier, various observations of star forming regions (Motte \& Andr\'e 1998, Andr\'e et al. 2007, Alfv\`es et al. 2007, Enoch et al. 2008)
confirm the similarity of the CMF and IMF, providing further evidence that stellar masses are directly linked to the core formation process. 
In this picture, at the heart of the present theory, the total (including further multiple systems) stellar masses are determined by the initial fragmentation of the cloud/clump into dense cores.
This is in contrast to alternative mechanisms where the stellar masses - thus the IMF - are dominantly determined by gas-to-star conversion
processes, such as competitive accretion, feedbacks or winds, for which the final IMF is not expected to reflect the CMF. 

Understanding the transformation of the prestellar {\it core} mass function (CMF), as described by the present theory, into the {\it stellar} (system) mass function (IMF) remains an open issue. The aforementioned observations suggest a nearly mass-independent conversion efficiency factor, in the $\sim 10$-50\% range, at least within the sensitivities of present surveys. Such a nearly constant star formation efficiency of cores is predicted theoretically if magnetically-driven outflows are the primary mediating factor from cores to stars (Matzner \& McKee 2000). Numerical experiments, on the other
hand, seem to suggest that the shape of the CMF is barely
affected by the nature of the core-to-star, i.e. CMF to IMF, conversion, except possibly in the very brown dwarf regime (Swift \& Williams 2008). Again, a more precise determination of the IMF
in this domain is needed to further explore this issue.

A substantial fraction of the prestellar cores, however, will fragment into multiple individual objects.
This sub-fragmentation episode very likely involves strong gravitational interactions and angular momentum redistribution
between the individual objects within a dense
core. Processes like radiative feedback, dynamical ejections, disk fragmentation, magnetic outflows might play some important role at this stage. An analytical
description of these strongly non-linear mechanisms seems to be rather elusive and we emphasize that the present theory only addresses the
initial stages of star formation, i.e. the formation of gravitationally bound prestellar cores. As such, this theory provides a sound theoretical foundation for the formation of pre-stellar/brown dwarf cores from the turbulent fragmentation of a cloud, and the related CMF/IMF. We also stress that, although relying on the same general concept
of gravo-turbulent fragmentation as the Padoan \& Nordlund (2002) theory, the present theory differs both qualitatively and quantitatively from the
Padoan-Nordlund one, with testable predictions, as discussed in \S7.2 of paper I.
Although the exact mass distribution derived in the present theory in the general (non-isothermal) case requires a (very simple) numerical resolution (eqn.(\ref{crit2})),
we provide in Appendix A a reasonably accurate {\it analytical} approximation of this CMF/IMF. This analytical expression should be used as a benchmark
to compare core/system mass distributions obtained from numerical simulations or inferred from observations, with the present theory. The HERSCHEL satellite, 
 launched in spring 2009, should reveal very-low-mass gravitationally bound prestellar cores and probe the CMF well into the substellar domain in various star forming regions. This will provide a stringent test for the mass distributions and the prestellar cloud conditions predicted by the present theory.



\acknowledgments
We thank Ralf Klessen for a critical reading of the manuscript and Enrique V\'azquez-Semadeni, the referee, for 
insightful comments.

This work was supported by the French "agence nationale pour la recherche (ANR)" within
the 'magnetic protostars and planets (MAPP)' project and by the "Constellation" european network MRTN-CT-2006-035890.

\appendix

\section{A simple and accurate approximation for the CMF/IMF}
As discussed in \S~\ref{crit_impl}, in the non isothermal case the 
criterion stated by eqn.~(\ref{crit2}) cannot be obtained analytically.
Since this makes the calculation of the CMF/IMF less straightforward, we propose
in this appendix an accurate and simple approximation.

In the purely thermal case, the criterion for collapse can easily be  obtained
for any value of $\gamma$ and is given by eqns.~(\ref{crit_therm}) and (\ref{crit2}), for $V_{rms}=0$ and ${\cal M}_*=0$. 
Our approximation consists to approximate the condition~(\ref{crit_therm}) by:
\begin{eqnarray}
M>M_J^{th}+M_J^{turb},
\end{eqnarray}
i.e. to assume that in order for the fluid cell to 
be gravitationally unstable, its mass must be greater than the sum of the 
critical masses corresponding to each individual support, thermal and turbulent.
In the case of a barotropic equation of state, as the one considered in 
\S~\ref{barotrop}, this gives for the critical mass:

\begin{eqnarray}
M \ge M_R^C  \simeq M_J^{0^\prime}(\gamma_1) 
\left( {R \over \lambda_J^{0^\prime} (\gamma_1)} \right)
^{{3\gamma_1-4 \over \gamma_1-2}} +
M_J^{0^\prime}(\gamma_2) \left( {R \over \lambda_J^{0^\prime} (\gamma_2)} \right)
^{{3\gamma_2-4 \over \gamma_2-2} } + 
M_J^{0^\prime}(\gamma_1) {\cal M}^2_*
  \left( {R \over \lambda_J^{0^\prime} (\gamma_1)} \right)
^{2 \eta + 1 }. 
\label{crit_approx}
\end{eqnarray}

\noindent This leads to:
\begin{eqnarray}
\tilde{M} \ge \tilde{M}_R^c \simeq {\tilde R}^{{3\gamma_1-4 \over \gamma_1-2}}  + 
  \kcrit^{1 \over 2 - \gamma_2} {\tilde R}^{{3\gamma_2-4 \over \gamma_2-2}} + 
 {\cal M}^2_*
  \tilde{R} ^{2 \eta + 1 }.
\label{crit_approx_norm}
\end{eqnarray}
where ${\tilde R}=R/\lambda_J^{0^\prime}(\gamma_1)$, 
 ${\tilde M} = M / M_J^{0^\prime}(\gamma_1)$ and $ \kcrit=(\rho^{crit}/ \bar{\rho} )^{\gamma_1-\gamma_2}$.
Thus 
\begin{eqnarray}
{d \tilde{M}_R^c \over d R} = {3\gamma_1-4 \over \gamma_1-2} {\tilde R}^{{3\gamma_1-4 \over \gamma_1-2}-1}  + 
   \kcrit^{1 \over 2 - \gamma_2} {3\gamma_2-4 \over \gamma_2-2} {\tilde R}^{{3\gamma_2-4 \over \gamma_2-2}-1} + 
 {\cal M}^2_* (2 \eta + 1)
  \tilde{R} ^{2 \eta }.
\label{crit_approx_deriv}
\end{eqnarray}
These quantities enter the final CMF/IMF, given by eqn.~(\ref{mass_spec}), that we rewrite here:
\begin{eqnarray}
{\cal N}(\widetilde{M}_R^c) = -{ \bar{\rho} \over M_J^{0^\prime}} {1 \over \widetilde{M}_R^c} {d \widetilde{R} \over d \widetilde{M}_R^c}
 {d\, \delta_R^c\over d\widetilde{R}}
 {1 \over \sqrt{2 \pi \sigma^2} } \exp \left( -{(\delta_R^c)^2 \over 2 \sigma^2} + 
{\delta _R^c \over 2} - {\sigma^2 \over 8}  \right), 
\label{mass_spec_app}
\end{eqnarray}
\begin{eqnarray}
\delta_R^c = \ln \left( { \widetilde{M}_R^c \over \widetilde{R}^3} \right), \\
\nonumber
{d\, \delta_R^c \over dR} =  {d \widetilde{M}_R^c \over d\tilde{R} } - {3 \over \tilde{R} }.
\label{dens_crit_app}
\end{eqnarray}
We also recall that 
\begin{eqnarray}
\nonumber
{\cal M}_* = { 1  \over \sqrt{3} } { V_0  \over C_s}\left({\lambda_J^{0^\prime} \over   1 {\rm pc} }\right) ^{ \eta}, \\\
\sigma^2 \simeq \ln( 1 + b^2 {\cal M}^2),
\label{Mstar}
\end{eqnarray}
where $\lambda_J^{0^\prime}$, given by eqn.~(\ref{lj0}), reflects the velocity dispersion at the Jeans length characteristic of the cloud's average density.
We now have an explicit set of equations straightforward to implement.
A comparison between this approximate expression and the exact one (solving eqns.~(\ref{crit_therm}) and (\ref{crit2})) for the second case displayed in fig.~\ref{fig_gamma_var}, 
 is shown in fig.~\ref{compar}.
As can be seen, the agreement is fairly reasonable, in particular the case $m=1$ (see eqn~(\ref{prescript}) for definition).

This approximated CMF/IMF should be sufficient for a first quick-and-easy comparison between the present theory and either observational or numerical results. More accurate comparisons, however, necessitate the proper resolution of eqn.~(\ref{crit2}), a fairly easy task.


\section{General formulation of the CMF valid for any equation of state}
In this appendix, we write the equations to be solved in the case of a
general equation of state written as: 
\begin{eqnarray}
{ C_s^2 \over  (C_s^0)^2} = f(\rho)=f( {\tilde{M_R^c} \over \tilde{R}^3}).
\label{eos}
\end{eqnarray}
Proceeding as before, we obtain:
\begin{eqnarray}
\tilde{M}_R^c = \tilde{R} \left( f(  {\tilde{M_R^c} \over \tilde{R}^3} ) + {\cal M}_*^2 \tilde{R}^{2 \eta} \right),
\label{eos}
\end{eqnarray}
leading to
\begin{eqnarray}
{ d \tilde{M}_R^c \over d \tilde{R}}  =
{ f(  {\tilde{M_R^c} \over \tilde{R}^3} )   - 3 {\tilde{M_R^c} \over \tilde{R}^3} f'({\tilde{M_R^c} \over \tilde{R}^3}) + (2 \eta +1) {\cal M}_*^2 \tilde{R}^{2 \eta} \over  1 - { 1 \over \tilde{R}^2 } f'(  {\tilde{M_R^c} \over \tilde{R}^3} ) }.
\label{der_eos}
\end{eqnarray}
which generalize eqns.~(\ref{crit_approx_norm}) and~(\ref{crit_approx_deriv}) while 
eqns.~(\ref{mass_spec_app})-(\ref{Mstar}) remain unchanged.
As described in \S~\ref{barotrop}, eqn.~(\ref{eos}) must be solved numerically.

\section{Core lifetime and the IMF}
The relationship between the core mass function and the initial mass function has been 
questioned by Clark et al. (2007) on the basis of a timescale argument.

Their point is the following: since the cores are dynamically collapsing, their typical timescale of evolution 
is the freefall time, $\tau_ {ff} \simeq 1/\sqrt{\rho}$. On the other hand, it is expected that the mass of the core
should be about one Jeans mass, $M \simeq M_J \propto C_s^3 / \sqrt{\rho}$. Thus, $M \propto \tau_{ff}$. This implies 
that while a core of mass M is collapsing, about 10 cores of mass $M/10$ should have time to form and collapse. 
Therefore, in order to produce an IMF having the slope of Salpeter, $x=1.35$, the CMF should have a slope $x=1.35-1=0.35$, not compatible 
with the observations.

We suggest that the solution to this problem relies on the fact that what determines the mass of the massive cores is not 
the thermal Jeans mass but the {\it turbulent Jeans mass}. Indeed, as discussed in paper I, this is the turbulent 
dispersion (taken into account in the parameter ${\cal M}_*$) which, in our theory, leads to the slope of Salpeter, $x =1.35$.
For the turbulent Jeans mass, we have $M_J^{turb} \propto V_{rms}^3 / \sqrt{\rho}$.

Thus, we have:

\begin{eqnarray}
M \propto { R^{3 \eta}  \over \sqrt{\rho} },
\label{mass_turb}
\end{eqnarray}
and since $M \propto \rho R^3$, 

\begin{eqnarray}
\rho \propto  M^{ -1 + \eta \over 1/2 + \eta}, \tau_{ff} \propto M ^{ 1 - \eta \over  1 + 2 \eta }. 
\label{mass_ff_turb}
\end{eqnarray}

Taking, for example $\eta=0.5$, we have $\tau_{ff} \propto M^{1/4}$ to be compared with the relation $\tau_{ff} \propto M$ obtained
by Clark et al. (2007). If, as it is the case in the present theory, the mass of the massive cores is determined by the turbulent 
dispersion, their density varies with their mass in a much  shallower way than if their mass would solely be determined by the thermal 
support. 

The implication is that the CMF has to be only slightly shallower than the IMF in order to obtain the Salpeter's slope. Recalling that 
the slope given by eq.~(\ref{x_salp}) is indeed slightly shallower than the Salpeter value, we believe that this is fully consistent with the 
core mass function being at the origin of the IMF. A more quantitative estimate including a full calculation is beyond the scope of this 
paper and will be considered elsewhere.

\end{document}